\documentclass[5p]{elsarticle}

\journal{Nuclear Fusion}
\usepackage{textgreek}
\usepackage{url}
\usepackage{epstopdf}
\usepackage{subcaption}
\usepackage{xcolor}
\usepackage{ulem}
\usepackage{tabularx}
\usepackage[modulo,switch]{lineno}
\usepackage{lipsum}
\usepackage{tabu}
\usepackage{booktabs}
\usepackage{subcaption}
\usepackage{amsmath}
\usepackage{float}

\makeatletter
\def\NAT@def@citea{\def\@citea{\NAT@separator}}
\makeatother

\begin{document}

\begin{frontmatter}

\title{Assessing the Risk of Proliferation via Fissile Material Breeding in ARC-class Fusion Power Plants}

\author[a]{John L. Ball}
\author[a]{Ethan E. Peterson}
\author[b]{R. Scott Kemp}
\author[a]{Sara E. Ferry}

\address[a]{Plasma Science and Fusion Center, Massachusetts Institute of Technology, Cambridge, MA 02139, USA}

\address[b]{Laboratory for Nuclear Security and Policy, Massachusetts Institute of Technology, Cambridge, MA 02139, USA}

\begin{abstract}
Construction of a nuclear weapon requires access to kilogram-scale quantities of fissile material, which can be bred from fertile material like U-238 and Th-232 via neutron capture. Future fusion power plants, with total neutron source rates in excess of $10^{20}$ n/s, could breed weapons-relevant quantities of fissile material on short timescales, posing a breakout proliferation risk. The ARC-class fusion reactor design is characterized by demountable high temperature superconducting magnets, a FLiBe liquid immersion blanket, and a relatively small size ($\sim$ 4 m major radius, $\sim$ 1 m minor radius) \cite{Sorbom_2015, kuang_arc_2018, Frank_2022}. We use the open-source Monte Carlo neutronics code OpenMC \cite{openmc} to perform self-consistent time-dependent simulations of a representative ARC-class blanket to assess the feasibility of a fissile breeding breakout scenario. We find that a significant quantity of fissile material can be bred in less than six months of full power operation for initial fertile inventories ranging from 5 to 50 metric tons, representing a non-negligible proliferation risk. We further study the feasibility of this scenario by examining other consequences of fissile breeding such as reduced tritium breeding ratio, extra heat from fission and decay heat, isotopic purity of bred material, and self-protection time of irradiated blanket material. We also examine the impact of Li-6 enrichment on fissile breeding and find that it substantially reduces breeding rate, motivating its use as a proliferation resistance tool. 

\end{abstract}

\begin{keyword}
ARC, proliferation, fissile, fertile, breeding, FPP
\end{keyword}

\end{frontmatter}

\section{Introduction}

In this work, we consider the proliferation risk associated with the operation of an ARC-class fusion power plant (FPP) in a breakout scenario. In the context of nuclear non-proliferation work, a ``breakout scenario" describes the deliberate misuse of nuclear technology intended for peaceful purposes for the production of weapons-usable material (WUM). Deuterium-tritium (D-T) fueled FPPs do not produce WUM under normal operating conditions. However, this does not imply that an FPP is inherently proliferation resistant. The D-T plasma is a high-flux, high-energy neutron source that can  be used to transmute non-weapons-usable fertile material into weapons-usable fissile material. Using OpenMC \cite{openmc}, an open-source radiation transport code, we analyze the potential for an ARC-class FPP to produce a weapons-relevant quantity of fissile material in less than one year of operation. This paper shows that it is theoretically possible to use an ARC-class plant to breed significant quantities of WUM in a short amount of time.

The purpose of this work is not to argue against the pursuit and adoption of fusion power. Nor is it to argue against the deployment of liquid breeder blanket concepts, which offer significant advantages with regards to tritium breeding and heat removal. Rather, we hope to show that breakout proliferation risk is a sufficiently serious concern for certain FPP concepts that it should be accounted for early in the design process. This is already the case for fission plants, which have an inherent proliferation risk due to their use of fertile and fissile material as fuel: in 2008, the U.S.\ Nuclear Regulatory Commission issued a Rule (a policy with the force of law) stating it expects fission reactor designers to consider diversion of weapons-usable material in the design phase \cite{73FR60612}. The fusion community should continue to analyze breakout risks associated with new designs, engage early with the relevant regulatory bodies to ensure that their operating plans are compliant with best practices and any fusion-applicable regulations, and proactively address public concerns. It is also important to devote resources to the development of technological strategies that can be incorporated into an FPP to achieve true proliferation resistance, including in the low-probability/high-risk breakout case.

First, we provide an overview of the basics of weapons proliferation and of prior work on the proliferation risk associated with fusion energy. We then discuss why the ARC-class tokamak is a particularly interesting case to study. Next, we detail the neutronics model used to complete this analysis. In the results section, we examine how adding between 5 and 50 metric tons of either natural uranium or thorium to the ARC-class FPP breeding blanket impacts the following  metrics: time to breed a significant quantity of weapons-usable material ($t_{\text{SQ}}$),  extra heat in the blanket due to fission, tritium breeding ratio (TBR), isotopic purity of produced fissile material, radiological hazards associated with the breeding of fissile material, and excess heat from radioisotope decay. We also consider how the isotopic enrichment of lithium in the blanket affects these parameters. Finally, we discuss the feasibility of this scenario in the context of the global nonproliferation regime and IAEA Safeguards, drawing upon historical examples for guidance.

The code and data that support the findings of this study are openly available at the following URL: \url{https://github.com/jlball/arc-nonproliferation}

\section{Background: nuclear weapons proliferation and fusion technology}

\subsection{Material needed to build a weapon}

Construction of a nuclear weapon requires access to kilogram-scale quantities of fissile isotopes like Pu-239 and U-233. These isotopes occur in low abundances in nature, so large quantities of raw material must be processed and isotopically enriched to build a weapon. An alternative path ``breeds" WUM from fertile isotopes in a neutron source (e.g. a fission reactor). Highly abundant, fertile isotopes like U-238 and Th-232 are transmuted into weapons-usable fissile isotopes via neutron capture:

\begin{equation}\label{eq:Un}
    ^{238}\textrm{U} + \textrm{n} \rightarrow \,^{239}\textrm{U} \xrightarrow[23.5~\textrm{min}]{\beta^-} \,^{239}\textrm{Np} \xrightarrow[2.356~\textrm{d}]{\beta^-} \,^{239}\textrm{Pu}
\end{equation}

\begin{equation}\label{eq:Thn}
    ^{232}\textrm{Th} + \textrm{n} \rightarrow \,^{233}\textrm{Th} \xrightarrow[21.8~\textrm{min}] {\beta^-} \,^{233}\textrm{Pa} \xrightarrow[26.98~\textrm{d}]{\beta^-} \,^{233}\textrm{U}
\end{equation}

\noindent The International Atomic Energy Agency (IAEA) defines a significant quantity (SQ) of fissile material as the approximate amount needed to make a first-generation weapon \cite{IAEA_safeguard_glossary}. For Pu-239 and U-233, one SQ is 8~kg or approximately $2 \times 10^{25}$ nuclei, and we use that definition in this work. Note that this definition has been criticized by some as being a factor of 2--3 higher than what is needed to make a nuclear weapon \cite{Goddard2017}. A substantial neutron source - such as a fission reactor - is therefore needed to transmute relevant quantities of fertile material on reasonable timescales. Commercial and researcher neutron generators have much lower source rates, ranging from $10^{7}$-$10^{11}$ neutrons per second, and are therefore incapable of breeding a SQ of fissile material on a timescale relevant to proliferation (months to years). Proposed commercial fusion reactors, however, are expected to have neutron source rates on the order of $10^{20}$ neutrons per second, and are therefore potentially useful for fissile material breeding.

\subsection{Proliferation resistance of D-T FPPs under normal operating scenarios}

Fission power plants pose an ``inherent'' proliferation risk because they contain and generate weapons-usable fissile material by default due to their fuel composition \cite{Jones2019}. Under normal operating scenarios, D-T fueled FPPs do not pose a significant proliferation risk because only a small amount of fertile or fissile material is expected to be present in the plant. Possible sources of fertile or fissile material on-site at an FPP include:
\begin{itemize}
\item The uranium coating inside fission chamber tubes, which may be used as a neutron diagnostic. This represents only a few grams of uranium and is not relevant to the production of significant quantities of weapons-usable material. 

\item Depleted uranium getter beds, which may be deployed in the fuel cycle system to capture hydrogenic isotopes. These getter beds represent small amounts of fertile uranium and are not relevant to the production of significant quantities of weapons-usable material. 

\item Uranium impurities that may be present in structural materials or breeder materials. It is possible that non-trivial amounts of fissile material could be produced in the blanket if adequate chemical purity standards are not implemented. A quantitative discussion of this issue as it pertains to FLiBe blankets is provided in Section \ref{subsec:impurity}.

\end{itemize}

While a normally operating D-T FPP will not generate significant amounts of WUM, it is important to note that it could still be used to support an existing nuclear weapons program through the diversion of Li-6 and/or tritium.Li-6 and tritium, in combination with deuterium, can enhance the performance of nuclear weapons.\footnote{In a fission implosion, D-T fuel provides additional neutrons which increase the yield of the fission device and reduce the amount and/or quality of fissile material needed, a process called boosting. In a two-stage thermonuclear device a large fraction of the total yield is from D-T fusion which is induced by the extreme conditions present after the detonation of a fission primary \cite{Goodwin2021}.}  An assessment of the risk of diversion of Li-6 and/or tritium from an ARC-class FPP is outside the scope of this work. The amount of tritium and Li-6 present on-site at an FPP will be highly dependent on both the design of a given plant's fuel cycle, the targeted TBR (which may be higher than the the TBR required for tritium self-sufficiency, pending operator decisions around safety margins) and the operational choices governing its tritium consumption and target tritium reserves.\footnote{We will note that research into FPP fuel cycle modeling shows that there will be limited excess tritium inventory available at D-T FPPs, suggesting that significant diversion of tritium produced in the blanket is likely to lead to a loss of tritium self-sufficiency in the plant \cite{abdou2020physics, meschini2023modeling}. This is true even considering that these models tend to make optimistic assumptions about the efficiencies of plasma operations, blanket breeding, and fuel cycle components, and do not account for inherent tritium losses due to trapping in components.} The fusion community has given significant attention to the topics of tritium and Li-6 diversion. For recent work on this topic, see \cite{diesendorf2023analyzing, Kovari_2018, PPPL_nonpro_workshop}.

\subsection{Prior work: fusion power and the breakout scenario}
\label{subsec:prior_work}

One category of relevant prior research focuses on fission-fusion hybrid plants. A hybrid plant consists of a fusion system coupled to a subcritical fission system such that some of the neutrons produced by fusion reactions interact with the fission section. The primary purpose of the hybrid design may be to produce power, to produce fissile material for use in conventional fission reactors from fertile material or spent fuel, and/or to burn waste actinides for easier disposal. In 1981, Conn et al.\ published a study on SOLASE-H, a fission-fusion hybrid designed to produce fuel for light water reactors \cite{conn1981fusion}. The authors considered how to ensure SOLASE-H would not be used for the production of WUM, and concluded that the intense radioactivity of the fuel assemblies would be a sufficient deterrent. Note that this concept presents an inherent proliferation risk, because it intentionally utilizes fertile and fissile material as fuel. Sahin et al. also considered fission-fusion hybrids in a series of papers from 1998--2001 \cite{Sahın_FusionBreeder_1998, Sahin_NeutronicAnalysis_1999, Sahin_NeutronicPerformance_2001}. In these works, fusion neutrons are used to breed U-233 from Th-232, and the fissile uranium is then used to fuel the fission plant. They propose fuel denaturing as a strategy to prevent the production of WUM, but this is not an option in breakout scenarios since the proliferator controls the isotopic profile of the fertile material. Vanderhaegen et al.\ considered a fission-fusion hybrid reactor in 2010 \cite{Vanderhaegen_Janssens-Maenhout_Peerani_Poucet_2010a} in which ThF$_4$ and UF$_4$ are dissolved in a FLiBe blanket attached to an ITER-class fusion device to breed fissile fuel for fission reactors. They concluded that LiF-BeF$_2$-ThF$_4$ was an ineffective choice for breeding fissile material, but that LiF-BeF$_2$-UF$_4$ resulted in relatively efficient production of fissile fuel. However, the LiF-BeF$_2$-UF$_4$ salt also produced a significant quantity of high-grade plutonium in a relatively short time ($<$~43 days), which presented a proliferation risk and made the concept less attractive. A 2012 conference paper by R.\ Moir also considered how fusion neutrons could be used to breed fissile material from thorium, and concluded that the radioactivity of the resulting material (due to an energetic gamma in the decay chain of co-produced U-232) would be a strong proliferation deterrent (see Sec.~\ref{subsec:purity} for further discussion of this issue).

Santarius et al.\ \cite{Santarius_Kulcinski_El-Guebaly_2003} considered whether passive proliferation resistance would be possible in a fusion power plant. Their conclusion was that the community should focus on the development of aneutronic fuel cycles if proliferation is considered to be a major risk. Realistically, the first FPPs will likely use the neutronic D-T fuel cycle as this reaction has the lowest requirements on plasma performance needed to achieve ignition. 

E.\ T.\ Cheng's 2005 paper \cite{Cheng_2005} explicitly looked at an FPP with a FLiBe tritium breeding blanket, as is also the case in this paper. In the Cheng study, the FPP is used as a neutron source to burn actinide waste from fission plants and/or produce additional fissile fuel from fertile material. This study also notes that (1) actinides are soluble in FLiBe and (2) can be removed from the FLiBe online, which are both points pertinent to this work. 

Sievert and Englert's 2010 paper \cite{sievert2010creating} considers the theoretical possibility of producing WUM from fertile material deliberately placed in the breeding blanket of an FPP, and indicates that proliferation safeguards will likely be needed for any future FPP that utilizes a tritium breeding blanket. Englert et al.\ considered the possibility of breeding plutonium in a Pb-Li blanket in greater detail \cite{Englert2010:1NMM10, Englert2011:INMM11} under scenarios in which homogeneous mixtures of natural uranium were introduced to the liquid Pb-Li in varying concentrations. These works used a simplified burnup model and suggested that the breeding of SQs of WUM in the blanket would certainly be possible, although the blanket design considered was highly simplified compared to more mature modern designs. Englert et al.\ noted that the fusion neutron spectrum and expected flux enabled the rapid production of WUM with very high isotopic purity, a finding that is consistent with the results of this work.

Glaser and Goldston performed an analysis of proliferation risks posed by magnetic fusion energy systems in 2012 that considered both clandestine and covert breeding scenarios \cite{glaser2012proliferation}. For the covert scenario, most pertinent to this work, the authors studied a DEMO class reactor with a Pb-Li blanket using the Monte-Carlo radiation transport code MCNP \cite{MCNP_6_3}. The authors suggest that fertile material could be covertly introduced into the blanket as TRISO particles to overcome the poor solubility of uranium and thorium in Pb-Li. They show that heat deposition from fission of fertile isotopes is substantial for uranium but not thorium, and that fusion power might have to be reduced to keep from overwhelming the plant's heat exchangers. The authors also show that the plant's TBR is negatively impacted by the introduction of fertile material to the blanket, with thorium being more detrimental than uranium. The authors also discuss how such a covert breeding scenario could be detected, including sampling of the blanket material and detection of characteristic gamma emission from fission products. It is concluded that a fast breeder fission reactor and a fusion reactor could produce WUM at similar speeds assuming they are of comparable power. The authors conclude that fusion systems present a lower proliferation risk than fission systems when appropriate IAEA Safeguards are implemented.

Franceschini et al.\ provide a useful overview of different schools of thought regarding fusion's risk within the non-proliferation community in their 2013 paper \cite{franceschini2013nuclear}. The authors point out that proposed fusion power plants would theoretically be able to produce SQs of WUM more quickly than fission reactors, using far less fertile material, and with lower radioactivity of the final product (leading to easier handling of the WUM). As fusion power is still years away, this is not treated as a major concern by the technical community: as long as fission power is on the grid, there is a tendency to assume that any nation intent on proliferation would rely on that technology first. The authors go on to outline the various technical, political, and regulatory conditions that would make fusion-enabled nuclear weapons proliferation more likely, on the assumption that fusion power eventually becomes a standard part of the energy generation mix. A 2023 paper by Diesendorf et al.\ further outlines the high-level risk scoping of proliferation hazards and the adequacy of existing safeguards in a ``mature fusion economy,'' in which fusion power is a widespread and common part of the energy mix, and concludes that a more rigorous risk assessment is merited \cite{diesendorf2023analyzing}.

\subsection{Weapons-usable material production and the ARC-class FPP}

This work focuses on the ARC-class FPP, a relatively recent design that is presently of significant interest to both the research and private fusion sectors. As explained below, the high-power-density and use of molten fluoride salt in the breeder blanket makes the ARC-class FPP a particularly interesting proliferation case study. ARC-class D-T FPPs were first proposed in 2015 \cite{Sorbom_2015}, with several subsequent design studies published since \cite{kuang_arc_2018, Frank_2022}. They are defined by the following broad characteristics: 
\begin{itemize}
\item The use of high-temperature superconducting (HTS) REBCO magnets with demountable joints to enable easy access to the interior of the plant for component maintenance and replacement
\item A compact, high-power-density design enabled by the high magnetic fields made accessible via HTS magnet technology
\item A replaceable vacuum vessel, which is adjacent to the first wall structure, and which is located \emph{inside} the tritium breeding blanket
\item A Liquid Immersion Blanket (LIB) for tritium breeding, currently assumed to use molten FLiBe salt as the breeding material
\end{itemize}
Prior to undertaking any quantitative analysis, we expected that an ARC-class FPP might present a breakout proliferation risk for the following reasons, and was therefore worthy of more detailed study: 
\begin{itemize}
\item Fertile material is known to be soluble in FLiBe. Fueled FLiBe has been studied experimentally and used practically. 
\item Actinide material is known to be removable from the FLiBe breeder via well-established processes.
\item The ARC-class FPP will have on-line and on-site capabilities that could be used for or modified to be capable of the addition, monitoring, and removal of actinide material. 
\item The ARC-class FPP has a high power density and high solid angle coverage of the neutron source with the breeder material, making it a potentially efficient breeder of WUM. 
\end{itemize}

The LIB is essentially a large volume of molten FLiBe (2LiF-BeF$_2$) salt that surrounds the first wall/vacuum vessel structure and is contained by a blanket-tank structure. This simple design is advantageous largely because it minimizes the amount of structural material needed for the blanket. This maximizes the amount of breeder volume exposed to neutrons (thus increasing the achievable TBR), enables more efficient heat transfer from the first wall/vacuum vessel structure into the blanket, and enhances shielding of the magnet structures as the low-Z elements of FLiBe are good neutron moderators. 

The use of FLiBe makes the ARC-class FPP particularly interesting from a fissile breeding proliferation standpoint, as it is well-established that fertile species are soluble in FLiBe. The most notable example of this fact is the Molten Salt Reactor Experiment (MSRE), which operated at Oak Ridge National Laboratory (ORNL) from 1965--1969 \cite{haubenreich1970experience}. The MSRE fuel was liquid LiF-BeF$_2$-ZrF$_4$-UF$_4$, with unfueled FLiBe as a secondary coolant. In the decades since, FLiBe has been extensively explored by the fission power research community as a fuel carrier for the general molten salt reactor (MSR) concept \cite{book_MSR_thorium, MSRs_and_IMSRs, Roper_2022}. 

In this work, we investigate what happens when fertile species (U-238 or Th-232) are introduced to the FLiBe LIB in amounts ranging from 5 to 50 metric tons, corresponding to a maximum molar percentage of 1.81\% and 1.84\% for UF$_4$ and ThF$_4$ respectively, assuming a blanket volume of 342~m$^3$ (see Sec.~\ref{subsec:modeloverview}). Upon introduction to the FLiBe, U and Th form LiF-BeF$_2$-UF$_4$ and LiF-BeF$_2$-ThF$_4$, respectively. Our upper limit is similar to fuel molar concentrations present in fission MSR designs \cite{MSBR_chem_rnd}.

In general, it is expected that an ARC-class FPP would already have on-line chemistry control capabilities, as well as on-site salt purification facilities. FLiBe with impurities and/or fuel compounds is known to be more corrosive to structural materials than pure FLiBe \cite{raiman2018aggregation}.\footnote{FLiBe chemistry and purity is of present interest to both the advanced fission community \cite{seifried2019general, zhang2019impurities} and the fusion community \cite{forsberg2020fusion}, with the latter particularly interested in how impurities could impact tritium breeding and extraction. Both are concerned with the long-term structural integrity of FLiBe-facing components.} These facilities might be capable of extracting bred WUM from the FLiBe during operation, significantly shortening $t_{\text{SQ}}$. If these facilities are undersized, they may at least furnish a prototype on which operators could construct a dedicated system capable of significant actinide extraction, either during operation or in a batch process afterwards.  

Removal of actinides from FLiBe is a well-established process. Fluorination of the fueled FLiBe converts UF$_4$ to gaseous UF$_6$, which then bubbles out and is collected \cite{FLiBe_Fluoride_Volatility, MSR_thorium_energy}. This technique was used to extract uranium from the MSRE fuel salt, but only a negligible amount of the plutonium present was extracted in this process as it did not volatilize \cite{MRSE_salt_processing}. However, it is possible to use the fluoride volatility process to convert PuF$_4$ to PuF$_6$ under the correct conditions \cite{FVM_Pu}.

\section{Methodology: OpenMC neutronics analysis}

The neutronics analysis for this work was carried out using OpenMC, an open-source Monte-Carlo radiation transport code \cite{openmc}. The code has been benchmarked against MCNP \cite{MCNP_6_3} and Shift \cite{Shift} for an ARC-class FPP model and shows good agreement \cite{bae_peterson_shimwell_2022}. The OpenMC depletion module, used extensively in the analysis presented below, has also been shown to agree well with FISPACT-II in fusion shutdown dose rate calculations \cite{Peterson_2024}. 

\subsection{Model Overview} \label{subsec:modeloverview}

\begin{figure}
    \centering
    \begin{subfigure}{\linewidth}
    \includegraphics[width=0.9\linewidth]{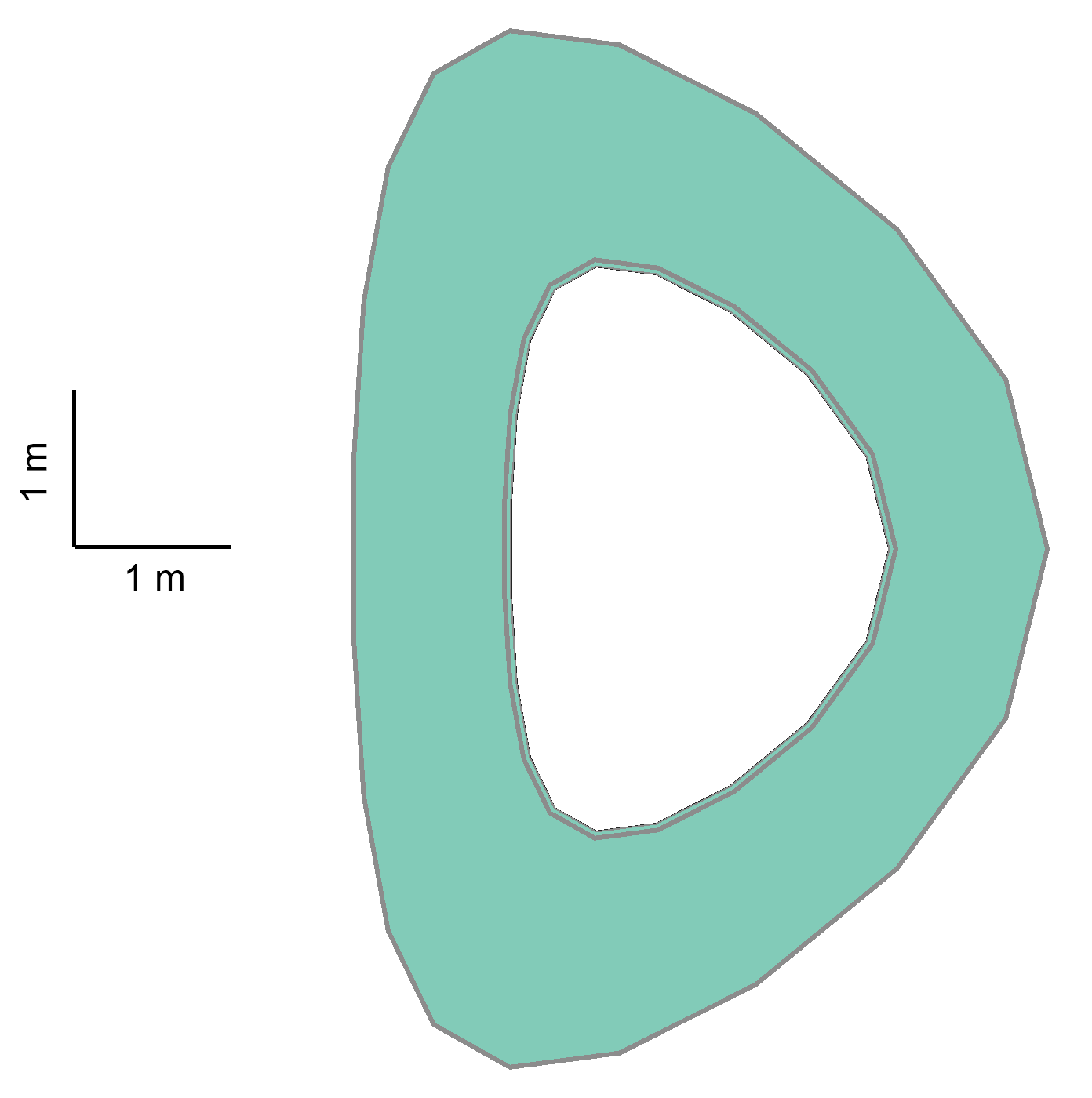}
\caption{\label{fig:poloidal}Plot of the poloidal cross section of the ARC-class FPP liquid immersion blanket studied in this work. The shape was generated using Eqs.~\ref{eq:R} and \ref{eq:Z} and the values in Tab.~\ref{tab:arcdesignpoint}.}
    \end{subfigure}

    \begin{subfigure}{\linewidth}
    \includegraphics[width=\linewidth]{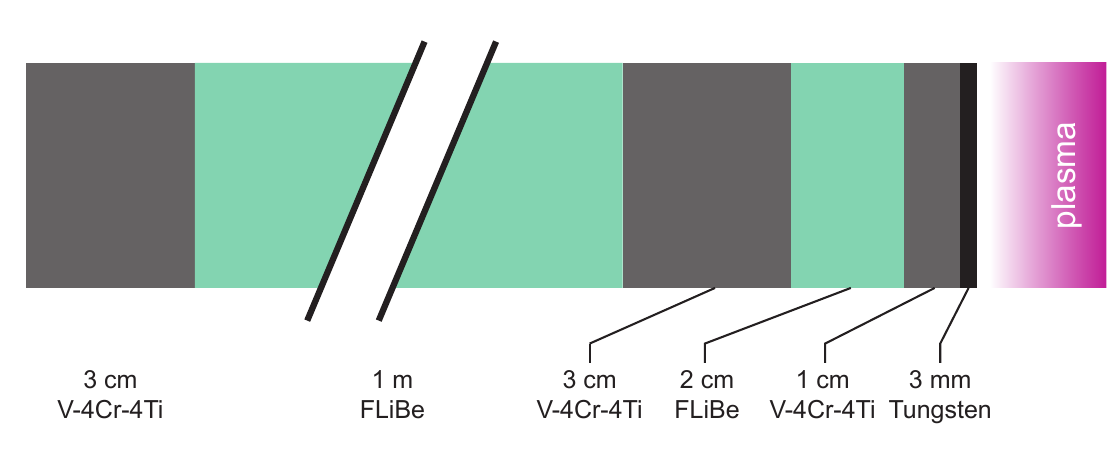}
    \caption{\label{fig:radial}Radial build of the OpenMC model of the ARC-class breeding blanket studied in this work.}
    \end{subfigure}

    \caption{The geometry and materials used to model the ARC-class FPP breeding zone in OpenMC for this work.}
    \label{fig:model_geo}
\end{figure}

First, we developed a representative geometric model of an ARC-class FPP tritium breeding blanket. Several ARC-class FPP design studies have been published, which we used to guide our chosen design point \cite{Sorbom_2015, kuang_arc_2018, Frank_2022}. Eqs.~\eqref{eq:R} and \eqref{eq:Z} describe the poloidal cross section shape of the blanket, which is plotted in Fig.~\ref{fig:poloidal}.

\begin{table}
\begin{center}
\begin{tabular}{|l|c|}
    \hline
    Major radius $R_0$ & 4~m \\ 
    Minor radius $a$ & 1~m \\
    Blanket thickness & 1~m \\
    Elongation $\kappa$ & 1.6 \\
    Triangularity $\delta$ & 0.5 \\
    \hline
\end{tabular}
\end{center}
\caption{\label{tab:arcdesignpoint} Plasma geometry parameters used in Eqs. \ref{eq:R} and \ref{eq:Z} to determine the plasma-facing component (PFC) contour.}
\end{table}

\begin{equation}
    R(t) = R_0 + a \cos{(t + \delta \sin{t})}
    \label{eq:R}
\end{equation}
\begin{equation}
    Z(t) = \kappa a \sin{t}
    \label{eq:Z}
\end{equation}

\noindent where $t$ parameterizes the RZ contour on the interval 0 to 2$\pi$, $R_0$ is the major radius of the machine, $a$ the minor radius, $\kappa$ the elongation, and $\delta$ the triangularity. Tab.~\ref{tab:arcdesignpoint} summarizes the values used to generate the OpenMC model. The radial build of the model is shown in Fig.~\ref{fig:radial}, and consists of six nested toroidal volumes representing the plasma facing components, vacuum vessel (with cooling channel), and blanket tank. Tungsten and  V-4Cr-4Ti were selected as the materials for the plasma facing components and vacuum vessel respectively. This RZ contour was then rotated  2$\pi$ radians about the major axis, forming a toroidally symmetric 360-degree model. Notably, this is a simplified model that does not include structures that would be present in a real FPP, such as RF heating and vacuum systems.

The neutron source is defined as a ring centered on the major axis of the device, emitting monoenergetic 14.1~MeV neutrons isotropically. We assume that the device is operated at a constant fusion power of 500~MW continuously in time for all analyses presented.

\subsection{Depletion calculation}\label{subsec:depletion}

We assume in this analysis that fissile material and fission products are not removed online and fertile material is not replenished online. It is thus necessary to consider transmutation and radioactive decay during irradiation, which is referred to as a depletion problem \cite{OpenMC_depletion}. We allow for the blanket composition to evolve in time by solving the Bateman equation \cite{Bateman_eq} numerically using the OpenMC depletion module \cite{OpenMC_depletion}, allowing the entire calculation to be carried out with just a single code. This is in contrast to codes like MCNP, which do not include a depletion solver and must be coupled to a second code like FISPACT-II \cite{FISPACT} to perform such a calculation. This, along with the fact that OpenMC is open-source (enabling the broader fusion community to more easily check these results, or adapt the source code to their own studies), motivates its use for this work. 

Notably, none of the prior work discussed in Sec.~\ref{subsec:prior_work} makes use of such a self-consistent simulation of fissile breeding. Approaching the calculation in this way is critical to ensure all effects relevant to breeding are resolved. The following phenomena are not captured with a time-independent method:

\begin{itemize}
\item Loss of bred fissile material to neutron reactions during breeding
\item Additional neutrons and heat from fission of bred fissile isotopes, which produce secondary neutrons and additional fission products
\item Neutron reactions on fission products
\item Radioactive decay of unstable nuclei, particularly the decay chains of fissile breeding reactions
\item Impurity-producing neutron reactions on intermediate daughter products in fissile breeding decay chains
\end{itemize}

The time stepping scheme was chosen in accordance with the accuracy criterion for the Chebyshev rational approximation method (CRAM), which is used by OpenMC to compute the matrix exponential for solving the Bateman equations. Since the primary neutron source is unaffected by reaction rates in the blanket, this is a conservative approach which minimizes errors in the methodology applied here.

\section{Results}\label{sec:results}

\subsection{Time to breed 1 SQ of weapons-usable material in an ARC-class FPP}
\label{subsec:tsq}

The critical parameter to determine the feasibility of breakout is $t_{\text{SQ}}$, the time required to produce one SQ of fissile material. As mentioned above, we assume that for the length of time these calculations represent, no fertile or fissile material is being removed by online chemistry control systems. This assumption minimizes the potential need for modifications to the FPP, but also slows the breeding process. Given that we simulate the system at a set of discrete time steps without knowing $t_{\text{SQ}}$ \textit{a priori}, we linearly extrapolate between the two time points with fissile masses just above and below a significant quantity to determine $t_{\text{SQ}}$. All results presented in this section assume a natural (unenriched) Li-6 content in the FLiBe of 7.5\%.

\begin{figure}
    \centering
    \includegraphics[scale=0.5]{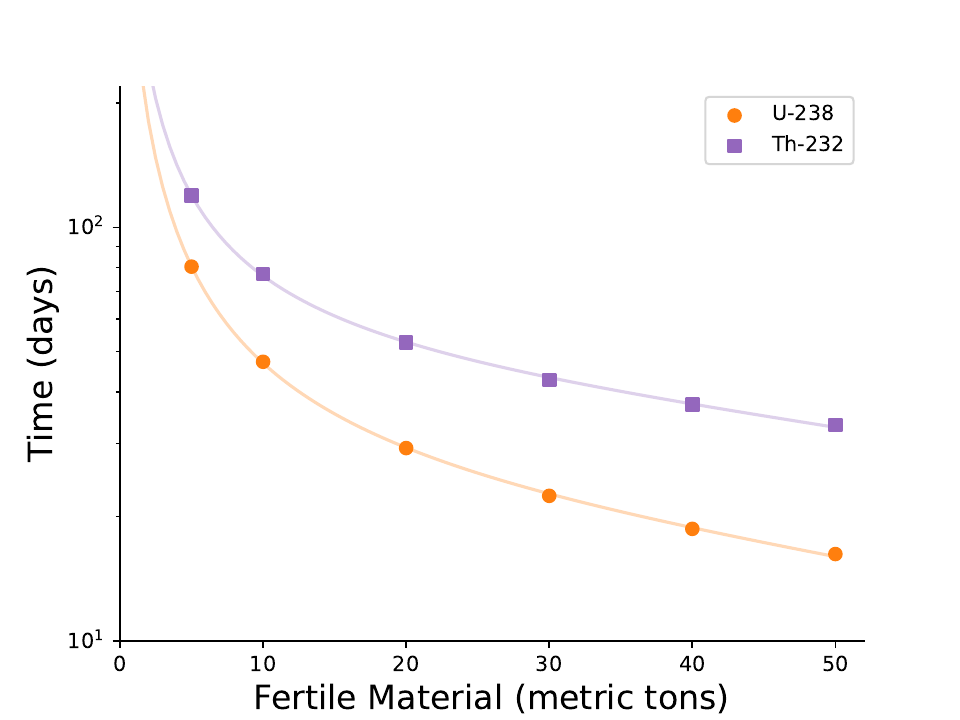}
    \caption{Plot of time to breed one significant quantity  ($t_{\text{SQ}}$)  of fissile material versus mass of fertile material dissolved in a representative ARC-class FPP liquid immersion blanket (see Sec.~\ref{subsec:modeloverview}) for two different fissile material breeding pathways (U-238 $\rightarrow$ Pu-239 and Th-232 $\rightarrow$ U-233; 1 SQ = 8~kg for both). We find that even for small amounts of fertile material input, ARC-class fusion reactors can produce 1 SQ of fissile material in less than 6 months, posing a possible proliferation risk.}
    \label{fig:time_to_sq}
\end{figure}

Fig.~\ref{fig:time_to_sq} plots $t_{\text{SQ}}$ as a function of fertile inventory initially dissolved in the blanket. Even for small quantities of fertile material ($\sim$2~metric tons), $t_{\text{SQ}}$ is less than one year for both the U and Th fuel cycles. Above 10 metric tons\footnote{Ten metric tons corresponds to about 1/2 cubic meter of U or Th metal, or one industry-standard 48Y-shipping container used for transporting natural UF$_6$.} of fertile mass, $t_{\text{SQ}}$ is on the order of 1-2 months for both fuel cycles.  For IAEA inspection purposes, it is important to note that the IAEA defines 10 metric tons as the SQ for natural uranium, and 20 metric tons as the SQ for natural thorium \cite{IAEA_safeguard_glossary}\footnote{This means that under current IAEA definitions, the diversion of less than 10 metric tons of natural U, or less than 20 metric tons of natural Th, over an inspection period would not trigger concerns of unsanctioned use of fertile material for weapons purposes.}.  These data can be fit using the following equation:

\begin{equation}
    t_{\text{SQ}}(m_f) = A/m_f - Bm_f + C
    \label{eq:fit}
\end{equation}

\noindent where $m_f$ is the mass of fertile material in metric tons, $t_{\text{SQ}}$ is in units of days, and $A, B,$ and $C$ are the fit coefficients. Their values are given in Tab.~\ref{tab:abc}. The fit described by Eq.~\eqref{eq:fit} is empirical and should not be assumed to apply for fertile masses far outside of the 5--50 metric ton range considered here. 

\begin{table}
\begin{center}
\begin{tabular}{|c|l|l|}
    \hline
    Fit Coef. & U-238 $\rightarrow$ Pu-239 & Th-232 $\rightarrow$ U-233 \\
    \hline
    $A$ & 330 $\pm$ 3.6 & 420 $\pm$ 8.2\\
    $B$ & 0.12 $\pm$ 0.014 & 0.25 $\pm$ 0.033 \\
    $C$ & 15.0 $\pm$ 0.62 & 37 $\pm$ 1.4 \\
    \hline
\end{tabular}
\end{center}
\caption{\label{tab:abc} Fit coefficients for Eq.~\eqref{eq:fit}, which is fit to the data plotted in Fig.~\ref{fig:time_to_sq}.}
\end{table}

We observe a significant difference in $t_{\text{SQ}}$ between the two production schemes, a result of the long half-life (26.98 days) of Pa-233 in the U-233 production chain, which creates a lag between the capture of a neutron by Th-232 and the actual appearance of U-233 in the blanket fluid. As 27~days is on the order of $t_{\text{SQ}}$ for all fertile mass inventories investigated, this result is expected. Note that the effect of Pa-233's long half-life is accurately captured using the self-consistent time-dependent depletion method, but would be lost if a simpler single-transport-calculation approach was used. We note however that this decay process does not need to happen within the FPP or while it is operation, thus the FPP could be shut down before the decay is completed, further reducing $t_{\text{SQ}}$ by reducing losses to fission reactions. 

\subsection{Fission heating in the blanket}

\begin{figure}
    \centering
    \includegraphics[scale=0.5]{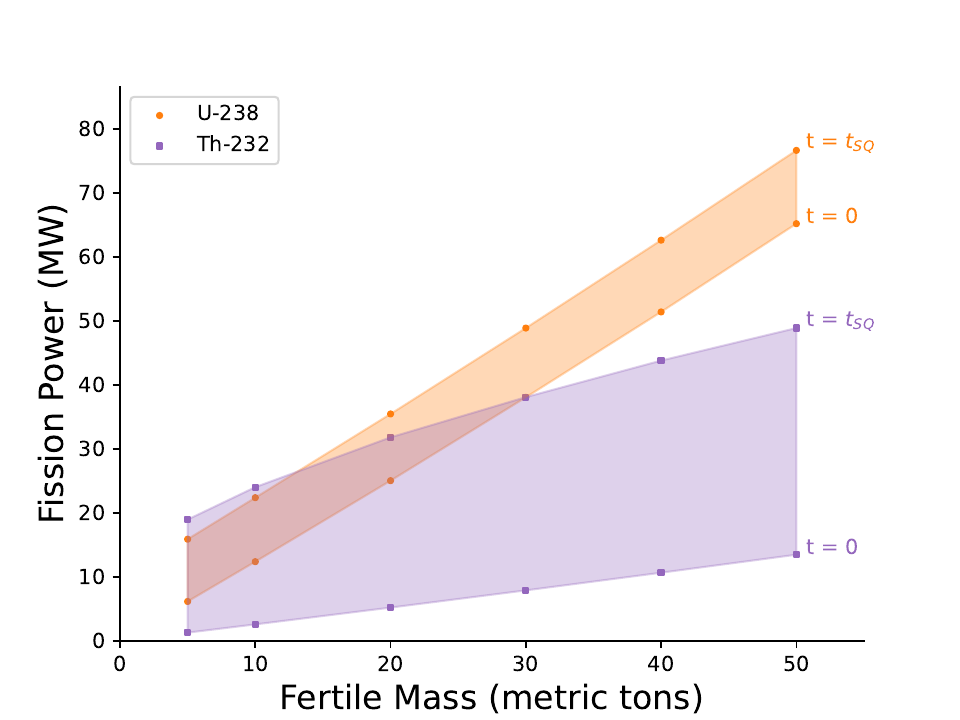}
    \caption{Plot of fission power in the blanket as a function of fertile inventory. The bottom curve of each colored region corresponds to fission at $t = 0$, when only fertile material is present, and the top curve corresponds to $t = t_{\text{SQ}}$, when 1 SQ of fissile material has been produced in the blanket. While we observe a substantial change in fission power as fissile material builds up in the blanket, fission power never exceeds $\approx$ 15\% of total fusion power (500MW), making it unlikely to limit a proliferation scenario.}
    \label{fig:fission_power}
\end{figure}

The introduction of fertile material into the tritium breeding blanket means that some rate of fission is expected. Fission reactions are exothermic and thus act as an additional source of heat in the blanket fluid. Fig.~\ref{fig:fission_power} plots fission power at $t = 0$ and $t = t_{\text{SQ}}$ as a function of fertile inventory in an ARC-class reactor blanket. For both production schemes, the fission power in the blanket over the plotted time interval is on the order of tens of megawatts. This represents a perturbation on the order of 10\% to 15\% of the total fusion power, assumed here to be 500 MW, indicating that excess fission power in the blanket is unlikely to make proliferation untenable unless safety margins on the heat exchanger components are very small. Even if this is the case, a proliferator could reduce the fusion power to create a total blanket heat load tolerable by the heat exchanger. Since the neutron rate is linearly correlated to fusion power, this reduction in source rate would be of the same magnitude as the heat from fission, shown here to be at most 15\% of nominal fusion power, increasing $t_{\text{SQ}}$ by the same amount to first order.

For Pu-239 production, fission of the U-238 fertile isotope is the dominant source of fission heating, with the build-up of Pu-239 resulting in only a small perturbation to the total fission power. For U-233 production, fission of the fertile isotopes is small compared to the fission power produced by bred U-233, as is shown by the large change in total fission power between $t = 0$ and $t = t_{\text{SQ}}$ in Fig. \ref{fig:fission_power}. 

\subsection{Impact of fertile material on tritium breeding} \label{subsec:tbrimpact}

The blanket of a D-T fueled FPP has three primary functions: it breeds tritium via interactions between the fusion neutrons and lithium in the blanket breeder material; it captures heat that is converted into useful energy; and it shields the magnets from neutron damage. The blanket's tritium breeding performance is characterized by the TBR, which is defined as the ratio of tritons produced via breeding to tritons consumed by fusion reactions in the plasma. For the reactor to be fuel self-sufficient, it must have a TBR $>$ 1, with additional margin that accounts for fuel cycle and fueling inefficiencies, tritium loss due to decay and uptake in materials, desired tritium inventory doubling time (for startup of new plants), and desired tritium reserve inventory. Therefore the required TBR is dependent on both plant design and operational decisions \cite{meschini2023modeling}. Generally, any scenario that results in a decrease of achievable TBR is detrimental to plant operations, and may result in a loss of tritium self-sufficiency (and thus an inability to continue operating the plant). 

\begin{figure}
    \centering
    \includegraphics[scale=0.5]{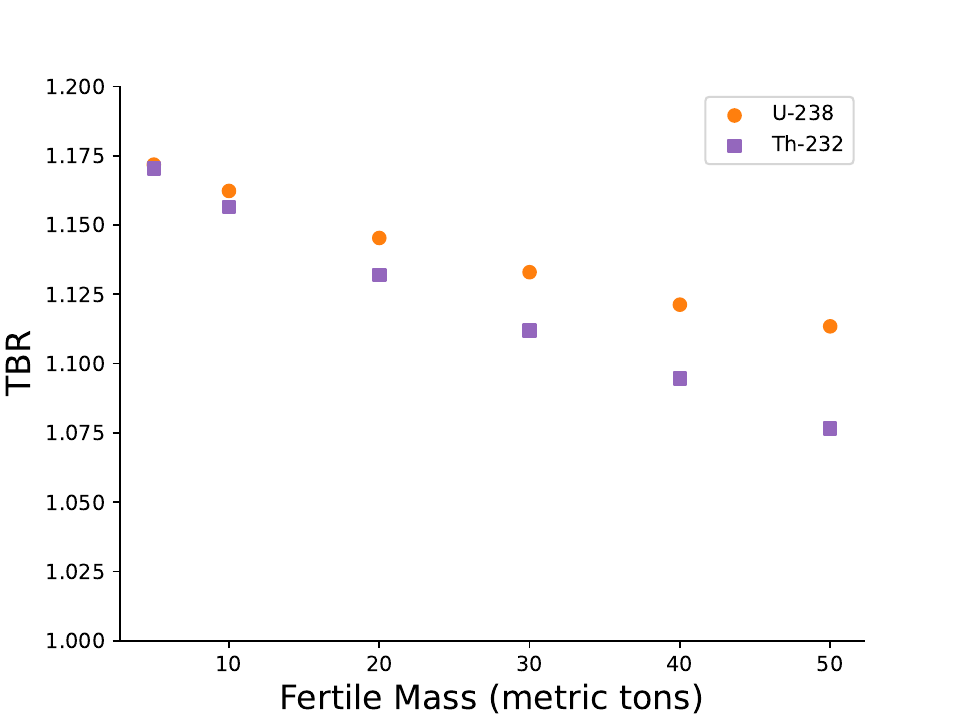}
    \caption{The tritium breeding ratio (TBR) in the model ARC-class FPP liquid immersion blanket is plotted as a function of fertile mass dissolved in the blanket for the fertile materials U-238 and Th-232. The presence of fertile material decreases the TBR, although not necessarily by an amount expected to render tritium self-sufficiency impossible. This determination depends heavily on the plant design and operational parameters.}
    \label{fig:tbr}
\end{figure}

The use of neutrons for breeding fissile material rather than tritium is expected to detrimentally impact TBR. We characterize the impact of introducing fertile material to the blanket on TBR in Fig.~\ref{fig:tbr}. TBR monotonically decreases with increasing fertile inventory, although TBR never falls below 1 for the plotted fertile mass range for either production scheme. Nonetheless, such a reduction may cause the TBR to fall below the level required for tritium self-sufficiency. In \cite{meschini2023modeling}, the TBR required for self-sufficiency in an example ARC-class FPP ranged from 1.012 to 1.113 for the parameters studied. Based on that assessment and the results in Fig.  \ref{fig:tbr}, TBR reduction due to the addition of fertile mass in the blanket would not necessarily be a proliferation deterrent. However, it is worth restating that our blanket model is simplified, and that the margin to loss of tritium self-sufficiency in a real FPP may be smaller. 

Note that Figure~\ref{fig:tbr} only shows data at $t = 0$, and does not account for the breeding of fissile material over time. In-blanket fissioning of fissile material boosts neutron flux inside the blanket which boosts TBR. However, this change was determined to be negligible and so time-dependent results were omitted from the plot.

\subsection{Isotopic purity of bred fissile material}
\label{subsec:purity}

\begin{figure}
    \centering
    \includegraphics[scale=0.5]{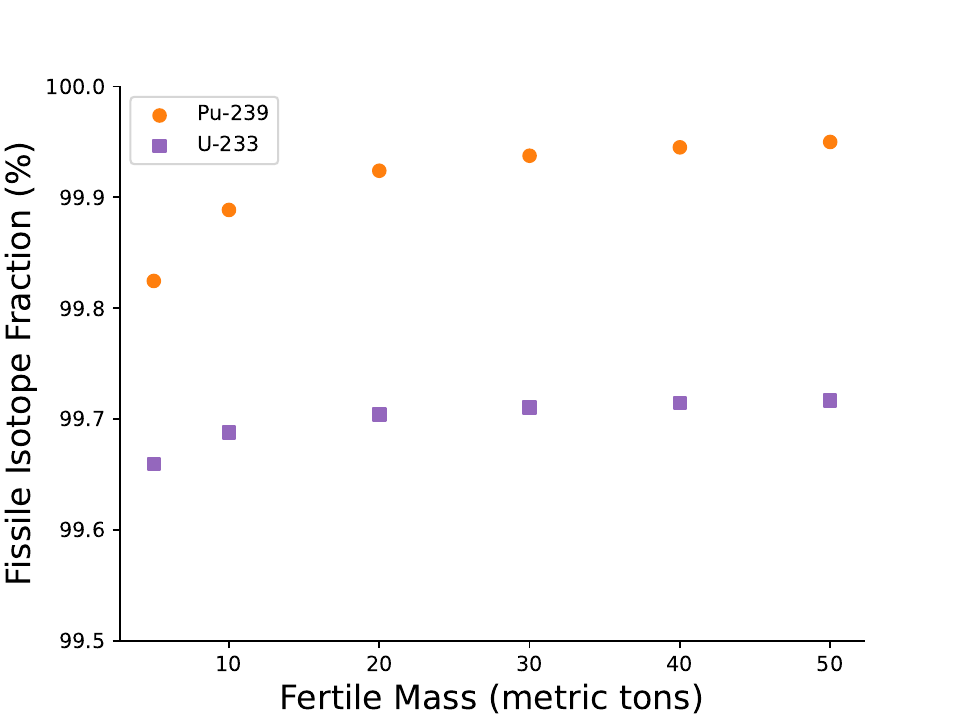}
    \caption{Plot of isotopic purity versus fertile inventory in the ARC-class FPP blanket. At all fertile masses considered, the isotopic purity achieved is well in excess of what is considered to be weapons-usable material.}
    \label{fig:isotopic_purity}
\end{figure}

Isotopic purity of bred fissile material is a key parameter in determining its usefulness as WUM. For example, Pu-239 can capture a neutron and become Pu-240, which is undesirable in a weapons context because of its high rate of spontaneous fission. Isotopic purity is computed at each time step by calculating the ratio of Pu-239 or U-233 nuclides to the total number of plutonium or uranium nuclei respectively. Fig.~\ref{fig:isotopic_purity} plots isotopic purity as a function of fertile inventory evaluated at $t_{\text{SQ}}$. We find that fertile inventory has little-to-no effect on the total isotopic purity at $t_{\text{SQ}}$, with $>$ 99\% purity for both production schemes. 

The purity of Pu-239 bred from U-238 is exceptionally high ($>99.8\%$ over the plotted fertile mass range), and well in excess of what is considered to be ``weapons-grade material'' ($>$93\% Pu-239 \cite{osti_plutonium_grade}). This is likely a result of the hardness of the neutron spectrum in the ARC-class FPP blanket, as the capture reactions that degrade isotopic purity are largest at lower neutron energies. Additionally, in this analysis we focus only on the breeding of a single SQ of WUM from metric tons of initial fertile inventory, representing a very low burnup fraction which is already well known in the fission community to correspond to high isotopic purity.

\begin{figure}
    \centering
    \includegraphics[width=\linewidth]{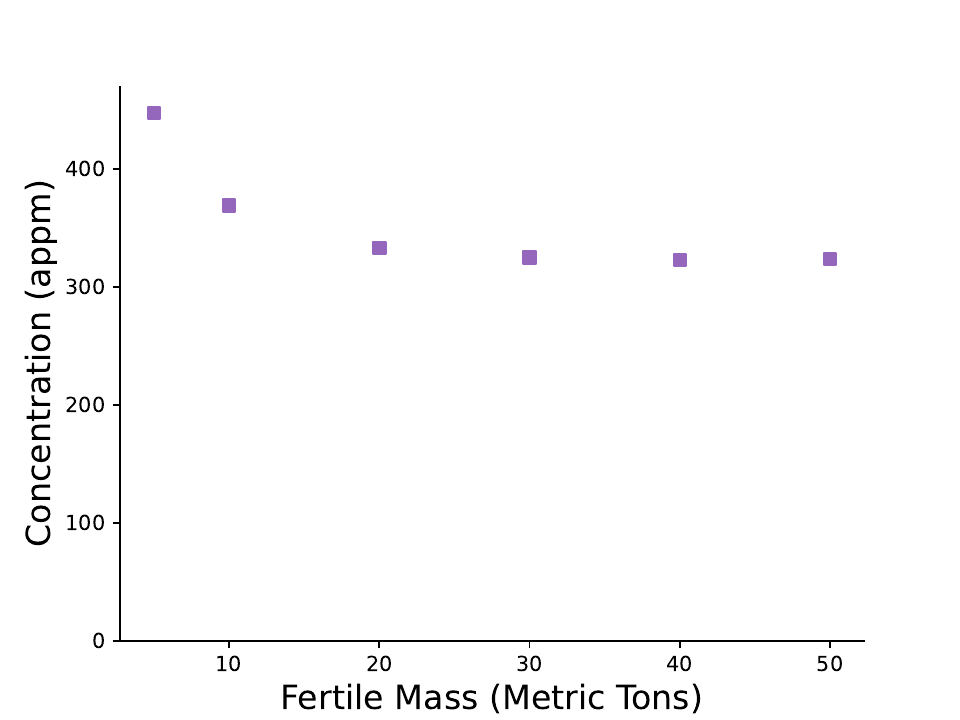}
    \caption{\label{fig:U232_content}Plot of U-232 impurity concentration in bred U-233 at $t = t_{\text{SQ}}$ in units of atomic parts per million (appm). The impurity U-232 has Tl-208 in its decay chain, a very active gamma emitter which creates substantial radiation hazards above 100 appm U-232, thus increasing the difficulty of manufacturing a weapon from contaminated material.}
\end{figure}

The purity of U-233 bred from Th-232 is also quite high ($>99.6\%$ over the plotted fertile mass range). However, it should be noted that the manufacture of a U-233 based weapon is complicated by the impurity U-232, which is co-produced with U-233 by mechanisms including radiative capture on Th-232 and (n,2n) reactions on the intermediate breeding daughter Pa-233 and U-233 itself \cite{U232_production}. The decay chain of U-232 includes Tl-208, which emits a 2.6~MeV gamma ray that makes working with contaminated U-233 very dangerous, complicating the manufacture of a weapon. Even small amounts of U-232 contamination, on the order of 100 ppm, can result in substantial radiation hazards. We find that for all initial fertile masses of Th-232, the concentration of U-232 at $t = t_{\text{SQ}}$ is very high ($>$300 ppm) if the U-233 is allowed to sit in the neutron flux until 1~SQ is obtained. Fig.~\ref{fig:U232_content} plots the concentration of U-232 in bred U-233 at $t = t_{\text{SQ}}$ for this scenario. The problem can be temporarily overcome via chemical processing to remove Tl-208 and other decay products, but dose rates will begin to rise again after a few weeks \cite{forsberg_1998_U233}. The problem can be more significantly overcome if online extraction of protactinium (the intermediate element in the transmutation of Th to U) from the salt is performed during the breeding process, but this requires a more complex modification to the salt purification system. 

\subsection{Self-protection time}
\label{subsec:self_protection}

Another consequence of fissile breeding is the production of fission products, which pose a radiological hazard as they decay. If sufficiently intense, this radiation could increase the difficulty of handling the contaminated salt, although this is certain to be much less problematic than handling fission-reactor spent fuel. We seek to determine if the radiation hazard posed by fission products would complicate the removal of the salt for reprocessing, thus increasing the probability of detection or extending the time needed to extract the bred WUM. 

NRC regulation 10 CFR \S 73.6(b) states that material with a dose rate greater than 1 Gy/hr at a distance of 1 meter without intermediate shielding is exempt from physical protection requirements as the radiological hazard is sufficient to prevent theft or diversion. In light of our simplified model, we use this rule to guide our calculation. We define the self-protection time to be the duration after shutdown at $t_{\text{SQ}}$ for which the dose rate 1 meter from the FLiBe is greater than 1 Sv/hr. We chose units of Sieverts instead of Grays as we assume that the dose is deposited in human tissue by gamma rays alone.

The self-protection time was computed by evolving the material composition at $t_{\text{SQ}}$ forward in time in the absence of neutron flux and computing the dose rate at each time step. To compute the dose rate, a simplified Monte Carlo model was used. An infinite 1 meter thick slab of actinide-doped FLiBe with gamma sources distributed uniformly throughout was modeled. The energy and activity of the gamma sources was based on the radionuclides present in the salt at that time step. 

\begin{figure}
    \centering
    \includegraphics[width=\linewidth]{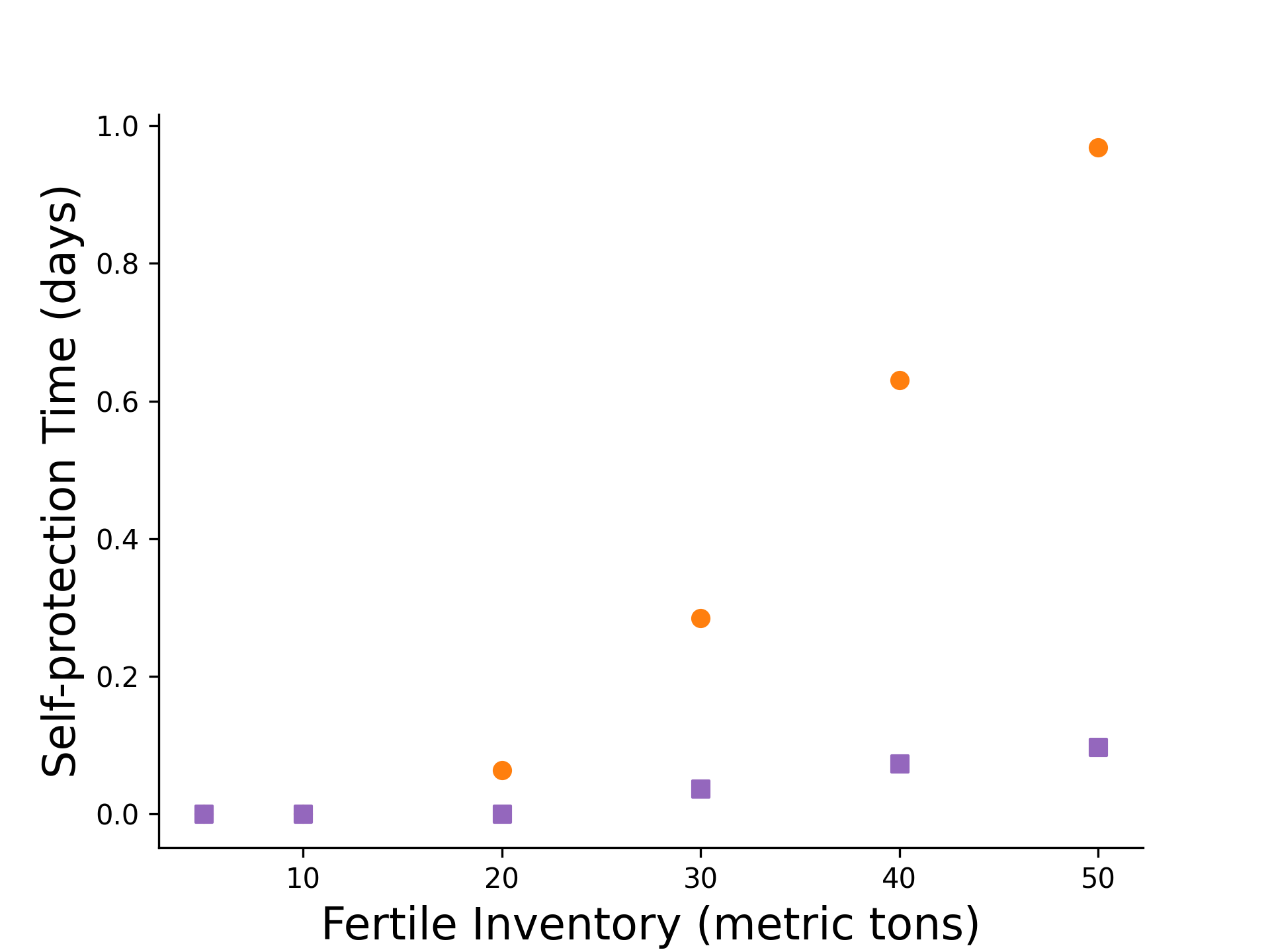}
    \caption{Plot of self-protection time, defined as the interval after shutdown for which the dose rate at 1 meter is greater than 1 Sv/hr, as a function of initial fertile mass. We find that if the reactor is shutdown at $t_{\text{SQ}}$ the self-protection time is at most a day for both breeding pathways and all mass inventories studied, and is thus not likely to prevent fissile breeding.}
    \label{fig:self_protection_time}
\end{figure}

Fig. \ref{fig:self_protection_time} plots self-protection time as a function of fertile mass. We observe that for all fertile mass inventories and both breeding pathways, the self protection time is at most one day. Given that the minimum $t_{\text{SQ}}$ determined above is 14 days, we find that self-protection time does not significantly increase the time to acquire fissile material from breeding nor will it impede its removal for reprocessing for an extended period of time. 

\subsection{Decay heat in the blanket}
\label{sec:decayheat}

\begin{figure}
    \centering
    \includegraphics[width=\linewidth]{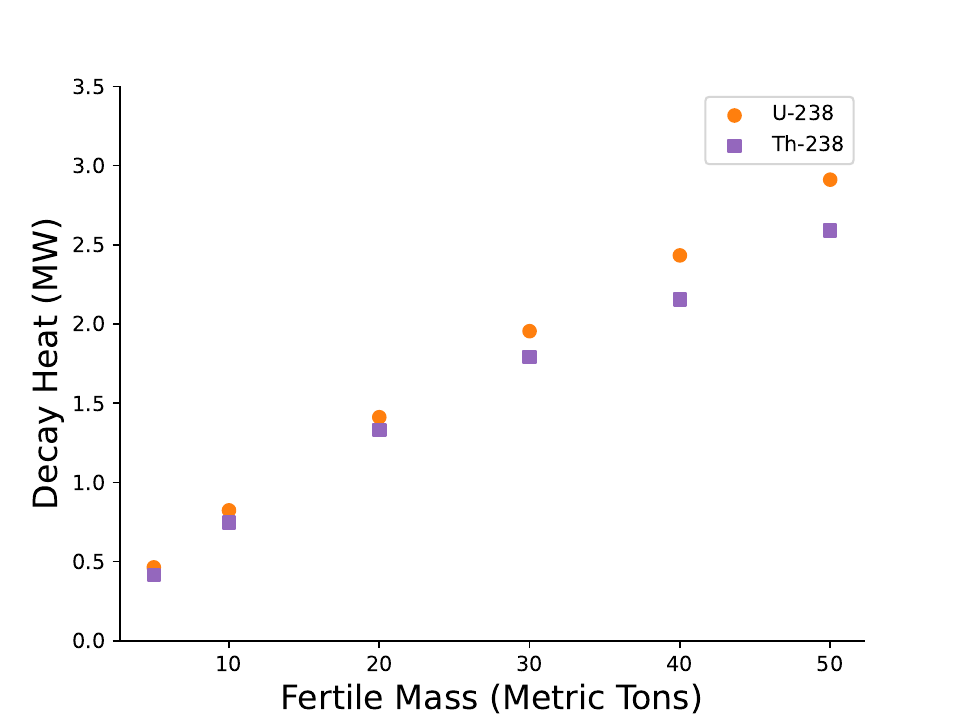}
    \caption{Plot of decay heat at $t = t_{\text{SQ}}$ versus initial inventory of fertile material, where $t_{\text{SQ}}$ is the time at which 1 SQ of fissile material exists in the blanket. Decay heat is a small ($<$5\%) fraction of total excess heat and represents a $<$1\% perturbation on total fusion power (500~MW). Thus decay heat is unlikely to limit the feasibility of a proliferation scenario.}
    \label{fig:decay_heat}
\end{figure}

Fissile material breeding in the blanket results in fissioning of some fissile elements and the subsequent presence of fission products. 
Radioactive decay of these products produces excess heat in the blanket.
Fig. \ref{fig:decay_heat} plots the decay heat in the blanket at $t = t_{\text{SQ}}$. Notably, the decay heat is a strong function of initial fertile mass, and peaks well below the maximum fission power observed. We find that at $t = t_{\text{SQ}}$ decay heat accounts for less than 5 percent of total excess blanket heating, and represents a $<$1\% perturbation to total fusion power. Thus, excess decay heat is not expected to be a significant deterrent to FPP operation in a breakout scenario. However if the salt is removed from the FPP for reprocessing, this extra heat will need to be removed. With volumetric heating densities of 1.5 - 9 kW/m$^3$ (depending on initial fertile inventory), heat removal for reprocessing is nontrivial. However, the decay heat magnitude decreases steadily:  in uranium-doped FLiBe, decay heat drops by an order of magnitude after $\sim$20 days, and in thorium-doped FLiBe the decay heat drops by an order of magnitude after $\sim$100 days.

\subsection{Impact of Li-6 enrichment}
\label{subsec:Li6_impact}

Li-6 enrichment has been widely considered as an option for improving the TBR of FPP blanket designs. This is because Li-6 has a large $1/v$ cross section for tritium breeding, while Li-7  has a non-zero tritium breeding cross-section only at very high neutron energies ($>$9 MeV). 

\begin{figure*}[p]
\centering
\begin{minipage}[b]{1\linewidth}   
\centering
\includegraphics[width=0.95\linewidth]{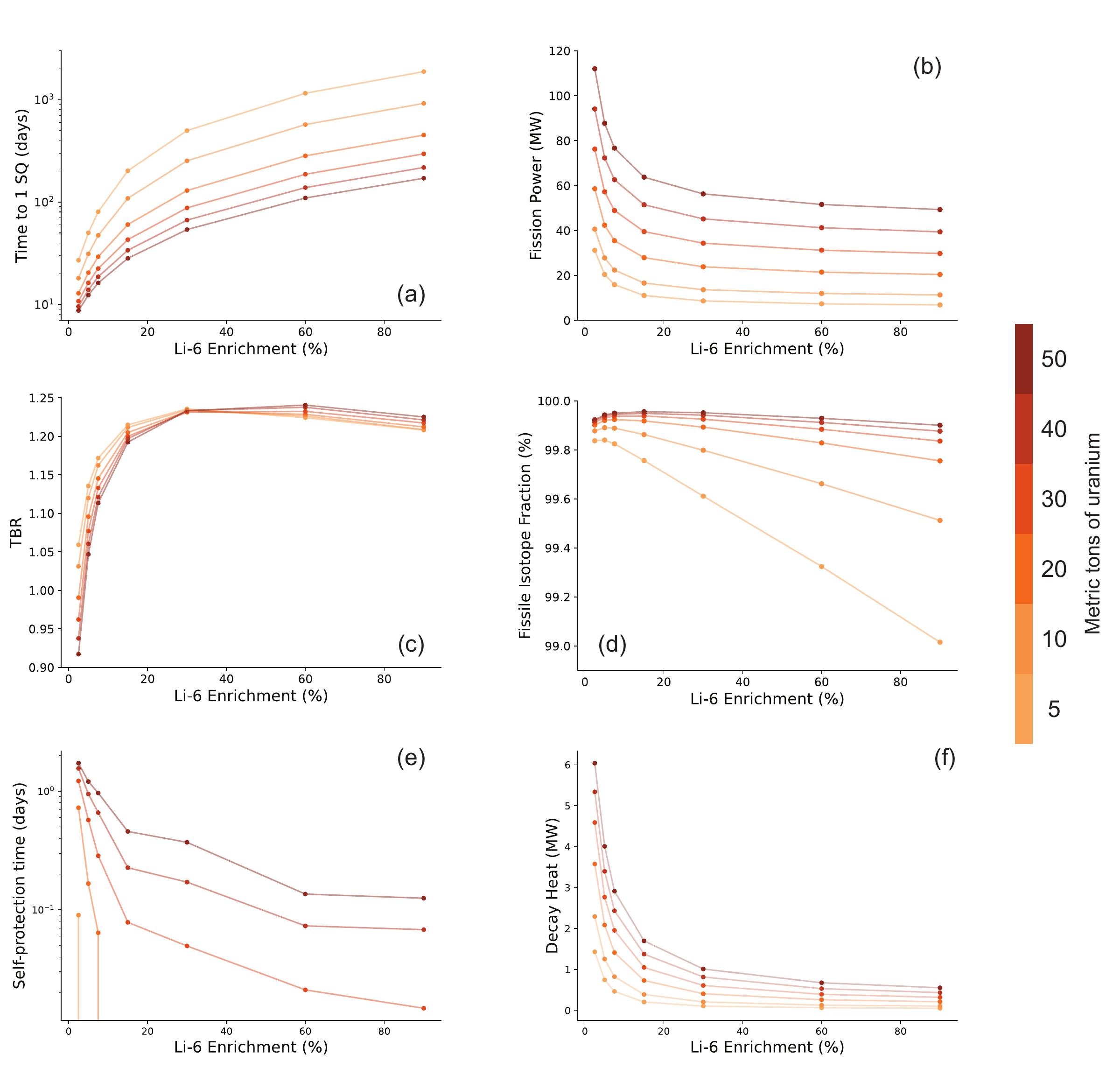}
\end{minipage}
\caption{\label{fig:U_Li_6} Plots of six relevant breeding parameters as a function of Li-6 enrichment for fertile mass inventories from 5--50 metric tons in the U-238 $\rightarrow$ Pu-239 production scheme. (a) Plot of time to 1 SQ vs.\ Li-6 enrichment. We observe a strong suppression of fissile breeding with increasing Li-6 enrichment, greatly increasing $t_{\text{SQ}}$ and motivating Li-6 enrichment as a proliferation resistance tool. (b) Plot of fission power at $t = t_{\text{SQ}}$ vs.\ Li-6 enrichment. We observe a reduction in fission power with Li-6 enrichment up to 30\% enrichment, after which fission power is approximately constant. (c) Plot of TBR vs.\ Li-6 enrichment. We observe that Li-6 enrichment increases TBR until $\approx$ 30\% enrichment, after which TBR is slightly reduced. However, the addition of fertile mass slightly increases TBR at  high Li-6 enrichment levels due to neutron multiplication reactions. (d) Plot of total isotopic purity vs.\ Li-6 enrichment. While we observe variations in purity with Li-6 enrichment, all purities remain in excess of 99\%, well above what is needed for use in a nuclear weapon. (e) Plot of self-protection time vs.\ Li-6 enrichment. We find that self-protection time decreases monotonically with Li-6 enrichment, which is expected given the increase in $t_{\text{SQ}}$ and reduction in fission rate. However, self-protection time never exceeds two days, and thus does not impact the viability of fissile breeding as a proliferation pathway. (e) Plot of decay heat at $t = t_{\text{SQ}}$ in the blanket material vs.\ Li-6 enrichment. We observe a monotonic reduction in decay heat with Li-6 enrichment. The magnitude of decay heats observed remains low ($\leq$ 1\%) across all Li-6 enrichments studied.}
\end{figure*}

\begin{figure*}[p]
\centering
\begin{minipage}[b]{1\linewidth}   
\centering
\includegraphics[width=0.95\linewidth]{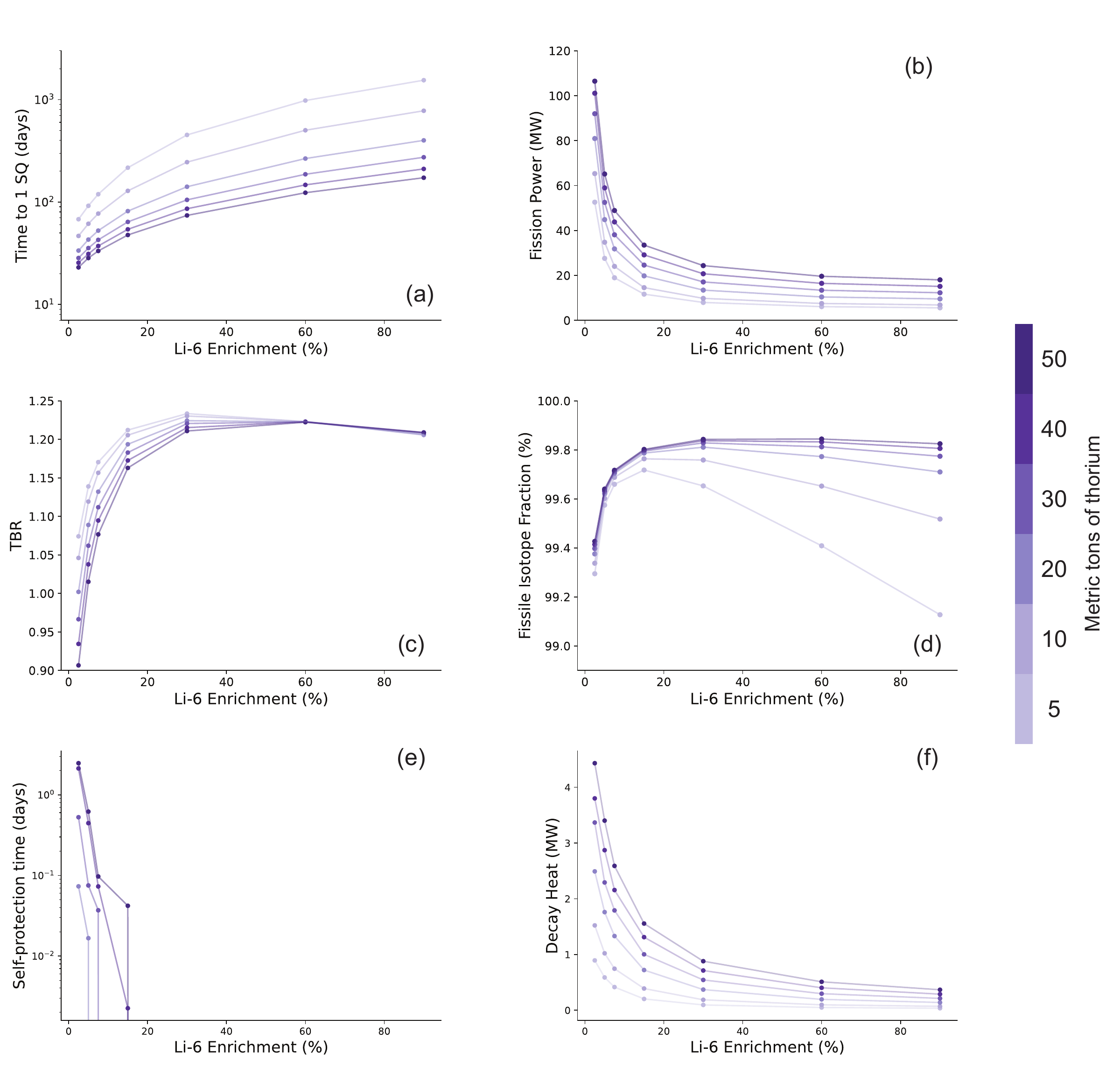}
\end{minipage}
\caption{\label{fig:Th_Li_6} Plots of six relevant breeding parameters as a function of Li-6 enrichment for fertile mass inventories from 5--50 metric tons in the Th-232 $\rightarrow$ U-233 production scheme. (a) Plot of time to 1 SQ vs.\ Li-6 enrichment. We observe a strong suppression of fissile breeding with increasing Li-6 enrichment, greatly increasing $t_{\text{SQ}}$ and motivating Li-6 enrichment as a proliferation resistance tool. (b) Plot of fission power at $t = t_{\text{SQ}}$ vs.\ Li-6 enrichment. We observe a reduction in fission power with Li-6 enrichment up to 30\% enrichment, after which fission power is approximately constant. (c) Plot of TBR vs.\ Li-6 enrichment. We observe that Li-6 enrichment increases TBR until $\sim$ 30\% enrichment, after which TBR is slightly reduced. Unlike in the U-238 $\rightarrow$ Pu-239 production scheme, no boost in TBR from neutron multiplication reactions is observed. (d) Plot of total isotopic purity vs.\ Li-6 enrichment. While we observe variations in purity with Li-6 enrichment, all purities remain in excess of 99\%, well above what is needed for use in a nuclear weapon. However, the U-233 is contaminated by very low levels of U-232, which produces highly radioactive Tl-208 as discussed in Sec.~\ref{subsec:impurity} and Fig.~\ref{fig:Li6_U232_content}. (e) Plot of self-protection time vs.\ Li-6 enrichment. We find that self-protection time decreases monotonically with Li-6 enrichment, which is expected given the increase in $t_{\text{SQ}}$ and reduction in fission rate. However, self-protection time never exceeds three days and for most cases is zero, and thus does not impact the viability of fissile breeding as a proliferation pathway. (e) Plot of decay heat at $t = t_{\text{SQ}}$ in the blanket material vs.\ Li-6 enrichment. We observe a monotonic reduction in decay heat with Li-6 enrichment. The magnitude of decay heats observed remains low ($\leq$ 1\%) across all Li-6 enrichments studied.}
\end{figure*}

We scanned Li-6 enrichment from  2.5\%  (below natural levels, which are $\approx$7.5\%) to 90\% enrichment to analyze its impact on the six breeding quantities of interest discussed in Sections \ref{subsec:tsq}-\ref{sec:decayheat} above. Overall results are presented in Figs.~\ref{fig:U_Li_6} and \ref{fig:Th_Li_6}.

\subsubsection{Li-6 enrichment and $t_{\text{SQ}}$}
Figs.~\ref{fig:U_Li_6}(a) and \ref{fig:Th_Li_6}(a) plot $t_{\text{SQ}}$ vs.\ Li-6 enrichment for the U-238 $\rightarrow$ Pu-239 and Th-232 $\rightarrow$ U-233 production schemes respectively. Li-6 enrichment is a strong lever on $t_{\text{SQ}}$, with 90\% enrichment increasing $t_{\text{SQ}}$ by about an order of magnitude over natural lithium for both production schemes. This creates a novel motivation for lithium enrichment in fusion systems as a tool for improving proliferation resistance as well as boosting blanket TBR. 

\subsubsection{Li-6 enrichment and fission heating} \label{subsec:li6fissionpower}

Figs.~\ref{fig:U_Li_6}(b) and \ref{fig:Th_Li_6}(b) plot fission power at $t = t_{\text{SQ}}$ in the blanket as a function of Li-6 enrichment. We observe that fission power is reduced with increasing Li-6 enrichment up to 30\% enrichment, after which there is little to no change. We observe this effect in both production schemes. This is likely a result of the hardening of the neutron spectrum with increasing Li-6 enrichment, which is discussed further in Sec. \ref{subsubsec:flux_spectrum}.

\subsubsection{Li-6 enrichment and TBR}

Figs.~\ref{fig:U_Li_6}(c) and \ref{fig:Th_Li_6}(c) plot TBR as a function of Li-6 enrichment. Unsurprisingly we observe that increasing Li-6 enrichment increases TBR, but only up to 30\% enrichment, after which there is little to no gain or even a slight reduction in both production schemes. In the U-238 $\rightarrow$ Pu-239 production scheme we also observe a slight boost in TBR with increasing fertile mass for enrichments above 30\%, likely the result of uranium's high rate of neutron multiplication and low probability of absorption in the faster spectrum created by Li-6 enrichment. For further discussion of Li-6 enrichment's impact on the neutron flux spectrum see Sec.~\ref{subsubsec:flux_spectrum}. We observe no such boost in TBR in the Th-232 $\rightarrow$ U-233 production scheme, and see an even greater reduction in TBR with fertile mass at lower enrichments, in line with the result of Sec. \ref{subsec:tbrimpact}.

\subsubsection{Li-6 enrichment and isotopic purity of bred fissile material}

Figs.~\ref{fig:U_Li_6}(d) and \ref{fig:Th_Li_6}(d) plot total isotopic purity of the bred WUM as a function of Li-6 enrichment. A maximum percentage of isotopic purity, more prominent for lower fertile mass inventories, is observed in both production schemes. However, at all conditions assessed here, the isotopic purities obtained are very high ($>$99\%) and more than sufficient for use in a nuclear weapon. Li-6 enrichment thus does not impact proliferation resistance from the standpoint of total isotopic purity.

\begin{figure}
    \centering
    \includegraphics[width=\linewidth]{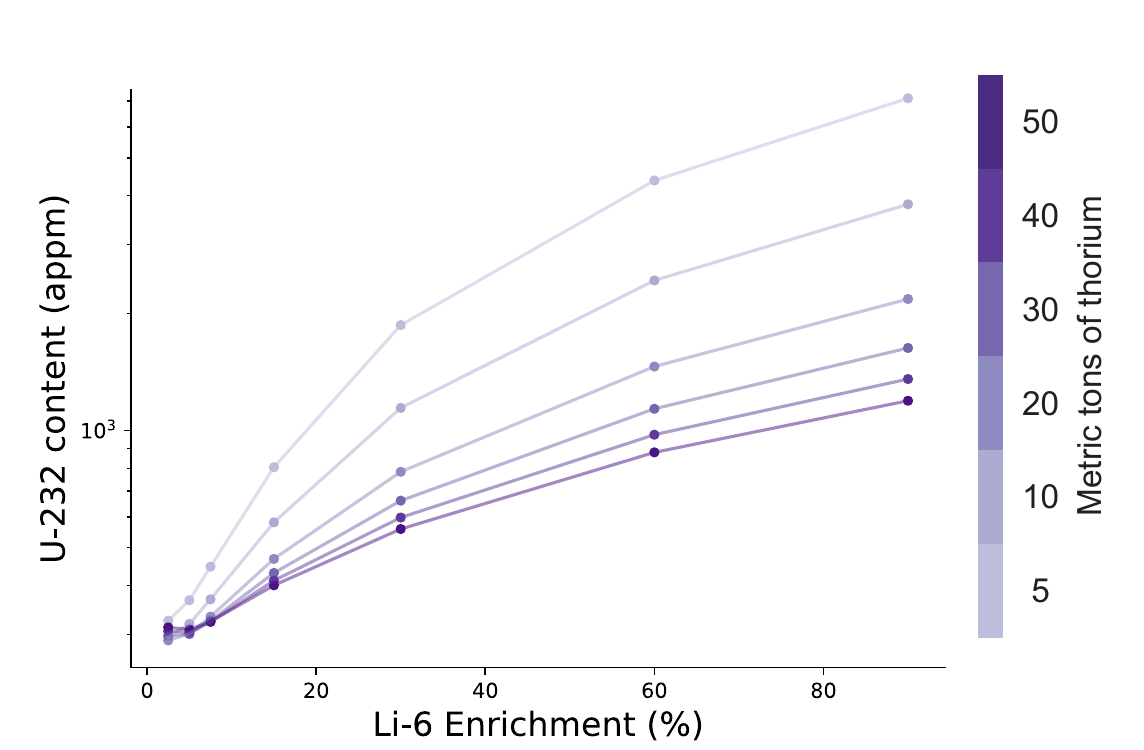}
    \caption{Plot of U-232 impurity concentration in bred U-233 (without online Pa removal) at $t = t_{\text{SQ}}$ as a function of Li-6 enrichment. We observe that Li-6 enrichment increases U-232 concentration for all fertile mass inventories, further motivating Li-6 enrichment as a proliferation resistance tool. See Sec.~\ref{subsec:impurity} for further discussion of the impact of U-232 on fissile material bred from Th-232.}
    \label{fig:Li6_U232_content}
\end{figure}

Fig.~\ref{fig:Li6_U232_content} plots the concentration of the U-232 impurity in U-233 at $t = t_{\text{SQ}}$ as a function of Li-6 enrichment. We find that Li-6 enrichment increases U-232 concentration for all fertile mass inventories studied, but the effect is larger for smaller inventories. Like the increase in $t_{\text{SQ}}$ resulting from Li-6 enrichment, this increase in U-232 impurity content further motivates Li-6 enrichment as a tool for proliferation resistance.

\subsubsection{Li-6 enrichment and self-protection time}

Figs.~\ref{fig:U_Li_6}(e) and \ref{fig:Th_Li_6}(e) plot self-protection time as a function of Li-6 enrichment. We observe that Li-6 enrichment monotonically decrease self-protection time for both production pathways, with notably all enrichments above 20\% having zero self-protection time for the U-233 production pathway. Given that self-protection time was already found to be very short for natural Li-6 enrichment, this result does not change the conclusion drawn in Sec. \ref{subsec:self_protection} that self-protection time does not substantially impact the feasibility of fissile breeding in ARC-class reactors.

\subsubsection{Li-6 enrichment and decay heat}

Figs.~\ref{fig:U_Li_6}(f) and \ref{fig:Th_Li_6}(f) plot the decay heat in the breeder material at $t = t_{\text{SQ}}$ as a function of Li-6 enrichment. We observe a monotonic decrease in decay heat with Li-6 enrichment, which is expected given the reduction in fission rate with increasing Li-6 enrichment. The difference between fertile masses varies widely over the range of Li-6 enrichments studied, with low Li-6 enrichments seeing much larger variations with fertile mass than higher Li-6 enrichments. The absolute magnitude remains small ($\approx$ 1\% of total fusion power) for all enrichments studied, indicating that decay heat is not likely to impact a proliferation scenario at any natural or greater Li-6 enrichment.

\subsubsection{Li-6 enrichment and the neutron flux spectrum in the blanket}
\label{subsubsec:flux_spectrum}

\begin{figure*}[ht]
\centering
\begin{minipage}[b]{1\linewidth}   
\centering
\includegraphics[width=0.95\linewidth]{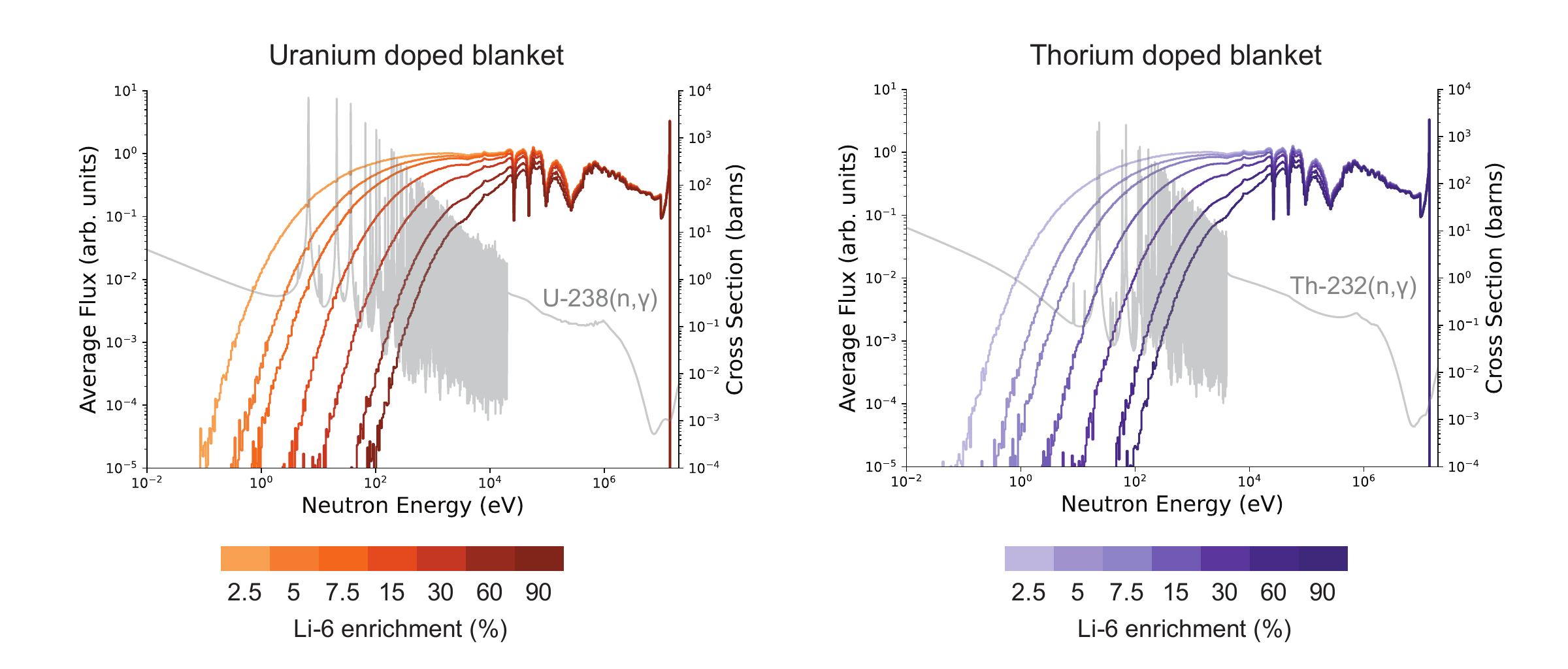}
\end{minipage}
\caption{\label{fig:Li_6_flux_spectrum}Plot of the neutron energy spectrum in an ARC-class FPP FLiBe blanket tank with 20 metric tons of fertile material as a function of Li-6 enrichment. Increasing Li-6 enrichment increases average neutron energy in the blanket, as fusion neutrons are more likely to be consumed in a Li-6(n,$\alpha$)T reaction before they are thermalized in the blanket and/or captured by fertile or fissile material via neutron capture and fission reactions. This increase in the average neutron energy with Li-6 enrichment is responsible for many of the trends, including suppression of fissile breeding, observed in Sec. \ref{subsec:Li6_impact}.}
\end{figure*}

To better understand the mechanism by which Li-6  impacts the fissile-material breeding process, we examine the average flux spectrum in the blanket tank as a function of Li-6 enrichment. Fig.~\ref{fig:Li_6_flux_spectrum} plots the average neutron energy flux spectrum in the blanket tank in arbitrary units for Li-6 enrichment ranging from 2.5--90\% for a 20 metric ton fertile inventory. We see a clear trend: increasing Li-6 enrichment significantly increases the average neutron energy. Fission and neutron-capture reactions in fertile and fissile material tend to have cross sections that peak at low neutron energies. These nuclear reactions are suppressed by the presence of Li-6, which tends to capture neutrons before they have time to be moderated to lower energies or interact with other nuclides. Note also that the Li-7 tritium breeding reaction produces neutrons, whereas the Li-6 tritium breeding reaction does not. Higher percentages of Li-7 in the FLiBe will contribute to the softening of the neutron spectrum by providing a source of lower energy neutrons.

\subsection{Localization of WUM breeding in the liquid immersion blanket}

\begin{figure}
    \centering
    \includegraphics[scale=0.15]{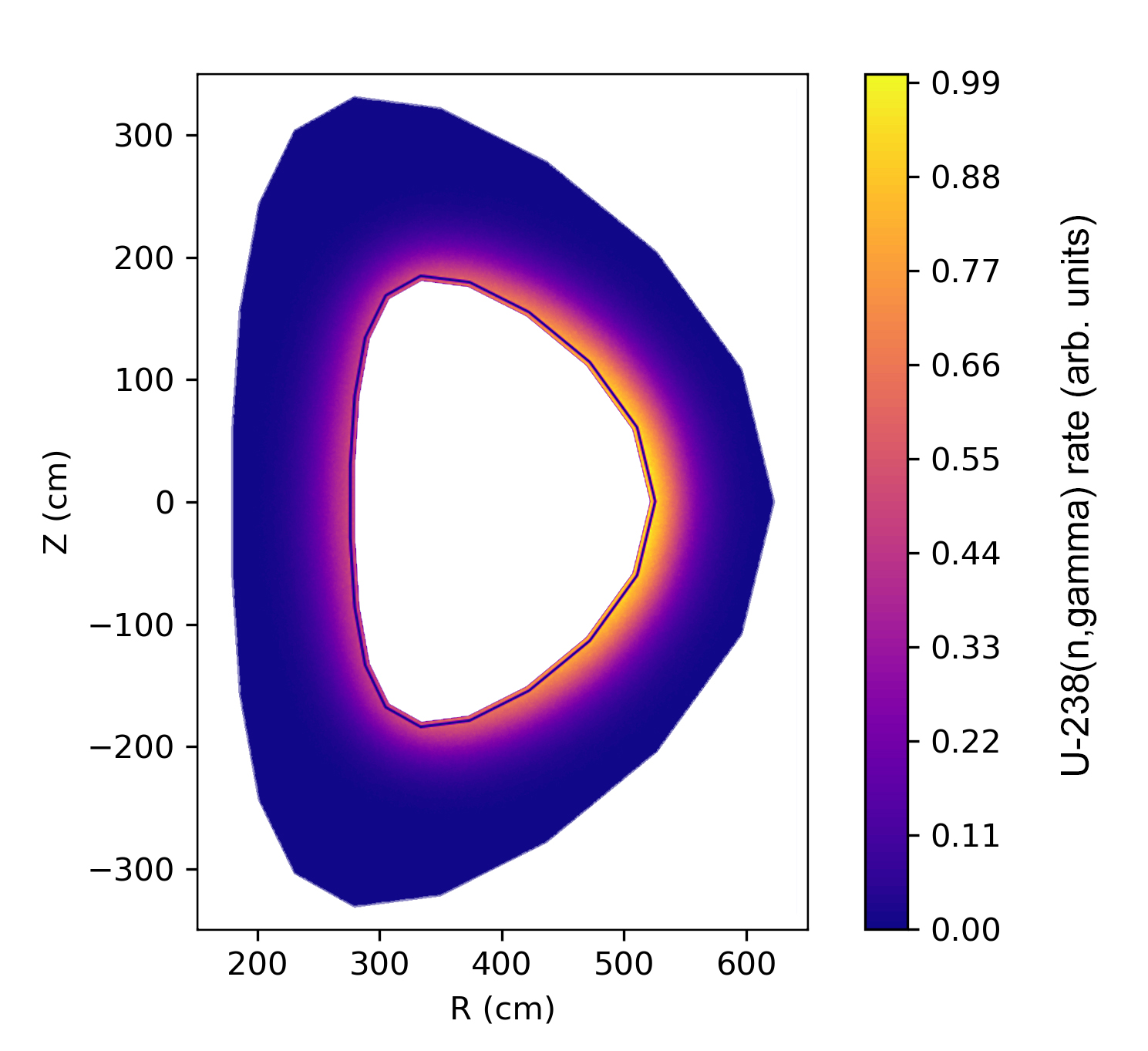}
    \caption{Tally of U-238(n,$\gamma$)Pu-239 reactions as a function of location in the blanket. The rate plotted is absolute and toroidally integrated, and is not volume normalized, but is normalized with respect to the maximum value obtained. Pu-239 production is highest closest to the neutron source (the plasma).}
    \label{fig:2d_breeding}
\end{figure}

It is sometimes assumed that fissile breeding is maximized further into the blanket, where the neutron spectrum is more thermal, as the relevant cross sections are highest at lower neutron energies. However, the OpenMC model indicates that WUM breeding is highest in the region closest to the plasma. This is visualized in Fig.~\ref{fig:2d_breeding}, which plots the reaction rate for U-238 neutron capture on a poloidal cross section of the liquid immersion blanket in the ARC-class FPP modeled in this work. Geometrically, the neutron flux is highest closer to the plasma neutron source (1/$r^3$ dependence); this effect outweighs the difference in cross section between higher and lower neutron energies.

This result shows that while our blanket model is simplified and slightly larger than a typical ARC-class device, we are completely capturing the relevant blanket regions for this phenomena and are not artificially introducing extra breeding volume. In fact, our use of a larger than necessary blanket is a conservative assumption in light of this result as the extra blanket material dilutes the dissolved fertile material, reducing the density of nuclides in the region of highest breeding and thus the breeding rate for a given fertile mass inventory. We note, however, that in a true system the actual volume of blanket material would be larger than the blanket tank volume alone, as some extra fluid must be present to be pumped through the heat exchanger and tritium extraction facilities. This reduces the conservativeness of this assumption.

\subsection{2D plots of relevant breeding parameters}

\begin{figure*}[ht]
\centering
\begin{minipage}[b]{1\linewidth}   
\centering
\includegraphics[width=\linewidth]{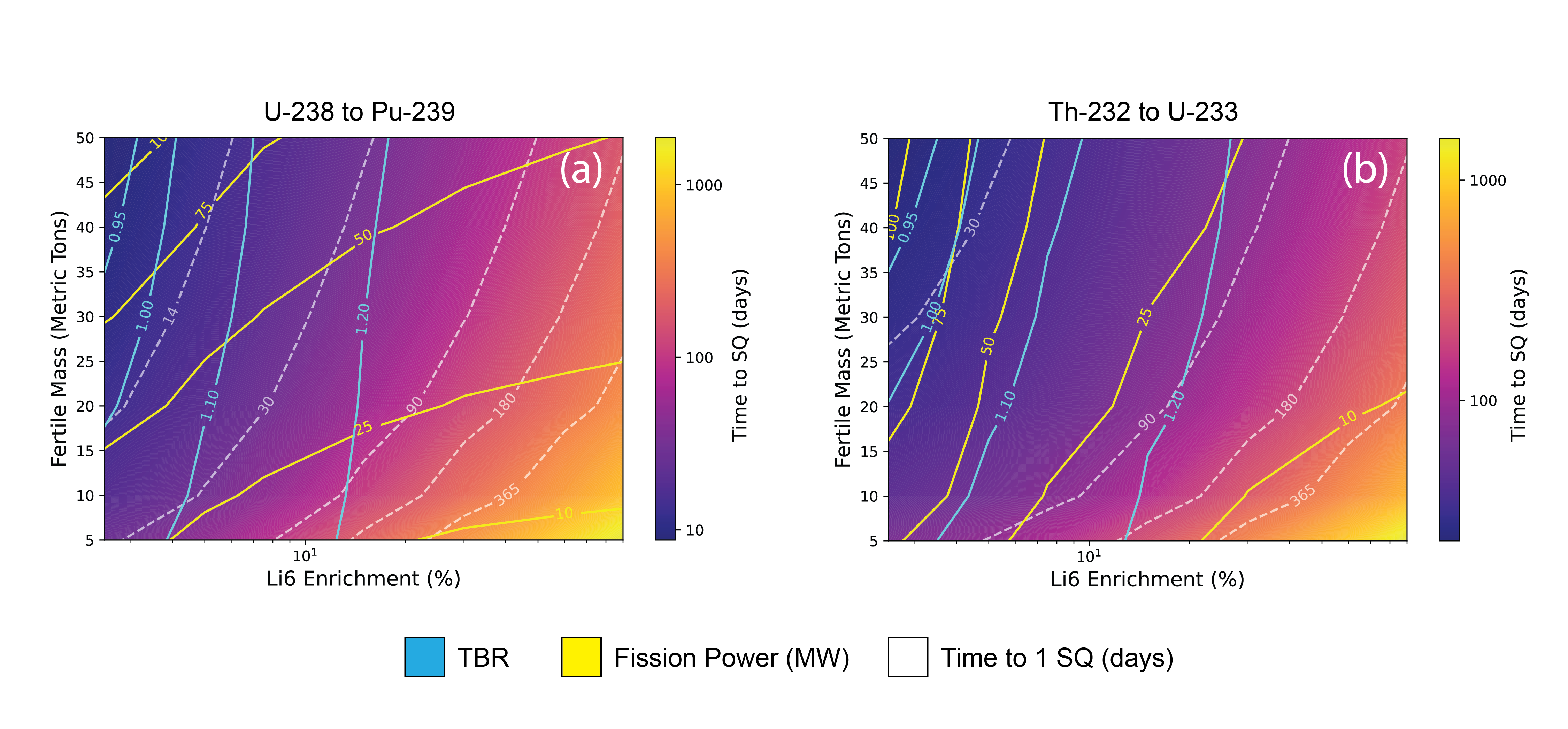}
\label{fig:StorageInvEvol_a}
\end{minipage}
\caption{Plots of three breeding-relevant parameters (time to 1 SQ ($t_{\text{SQ}}$), TBR, and fission power) as a function of Li-6 enrichment and fertile mass in the blanket for both the (a) U-238 $\rightarrow$ Pu-239 and (b) Th-232 $\rightarrow$ U-233 production schemes. These plots summarize some of the key results presented in Figs.~\ref{fig:U_Li_6} and \ref{fig:Th_Li_6} in a way that allows for easier comparison across parameters.}
\label{fig:proliferation_plot}
\end{figure*}

So far, we have examined the dependence of $t_{\text{SQ}}$, fission power in the blanket, isotopic purity of bred WUM, TBR, decay heat, and self-protection time on fertile mass and Li-6 enrichment. Fig.~\ref{fig:proliferation_plot}(a) and (b) present this data in one plot for both production schemes, allowing quick identification of the 2D parameter regions that are most concerning for proliferation. We have opted not to include decay heat, isotopic purity, or self-protection time as these quantities have been shown to have no substantive impact on the feasibility of fissile breeding anywhere in the parameter space studied.

These plots summarize the key results of this work. Enriching the FLiBe blanket in Li-6 has dual benefits of improving TBR (the usual motivation for lithium enrichment) and significantly increasing $t_{\text{SQ}}$. Importantly, a region where $t_{\text{SQ}}$ is $>$ 1~yr appears only for Li-6 enrichment $>$ 20\%. At 90\% enrichment, this region covers about half of the fertile masses studied. Thus a large portion of the 2D parameter space visualized is of possible proliferation concern. While in this work we are unable to definitively set limits on quantities like TBR and fission power which might limit a proliferation scenario, we hope that future more detailed reactor design studies could make use of similar plots to analyze their vulnerability to this proliferation pathway.

\subsection{Uranium impurities in beryllium and implications for plutonium production} \label{subsec:impurity}

Natural beryllium can be significantly contaminated with impurities, including uranium. Therefore, it is well known that fusion concepts using significant amounts of beryllium as a neutron multiplier should consider the effects of uranium contamination, including plutonium production, and consider standards for beryllium purity. ITER research teams in particular have given this matter significant attention as beryllium was long-intended to be used as a first wall material in the device (although this changed in 2023). One calculation determined that Pu-239 production in ITER would be on the order of a gram after five years of operation (for 1 wppm of U impurity in beryllium), but could be on the order of 10 kg in DEMO-class plants \cite{Cambi_Cepraga_Di_Pace_Druyts_Massaut_2010}. 

Here, we consider how uranium impurities in beryllium could result in plutonium production in the ARC-class FPP under normal operating conditions for uranium impurity levels of 50, 100, 150, and 200 weight parts per million (wppm), representing 2.4, 4.8, 7.2, and 9.6 kilograms of fertile mass, respectively. To put these values into context, Materion specifies its S-65 grade of beryllium as having a maximum of 150 wppm of U impurity \cite{materionBe}. Beryllium mined from Russia or Kazakhstan were found to have an average of 5.2 wppm uranium impurities (with different Be samples ranging from 0.16 to 18 wppm U content) \cite{kolbasov2016use}. Beryllium obtained for use in the Advanced Test Reactor was determined to have an average uranium impurity content of 71 wppm (with concentrations ranging from 23--105 wppm) \cite{longhurst2007}. 

\begin{figure}
    \centering
    \includegraphics[width=\linewidth]{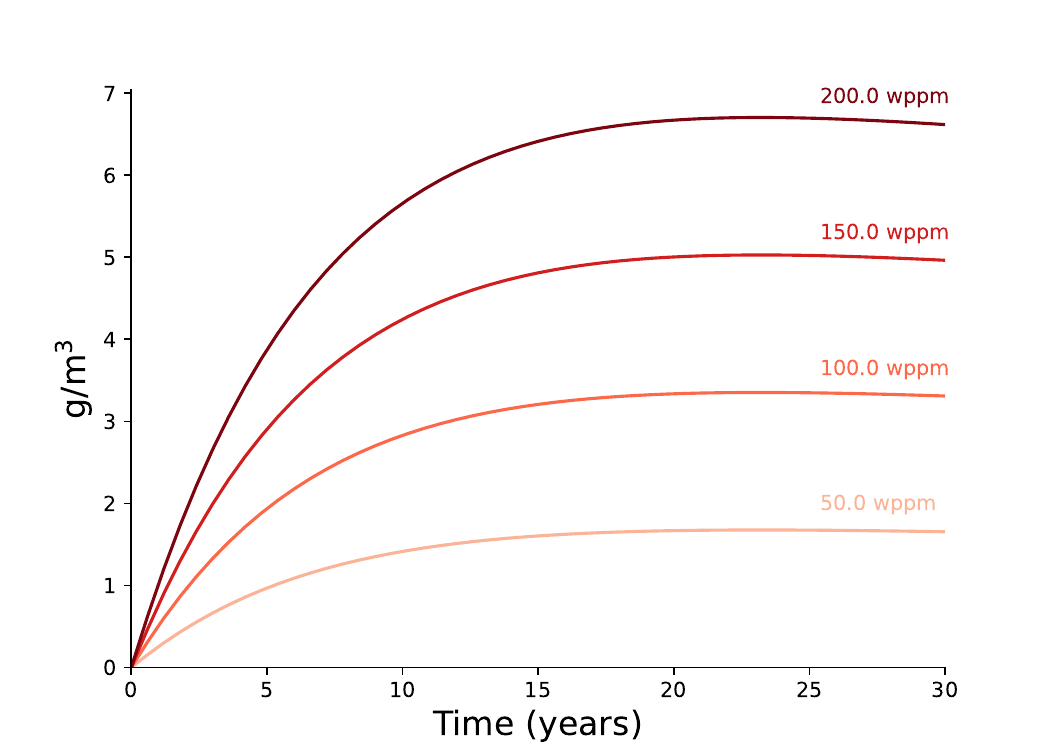}
    \caption{\label{fig:impurity_fissile_mass} Plot of Pu-239 mass density accumulated in the ARC-class FPP liquid immersion blanket for varying levels of naturally occurring uranium impurities in the beryllium of the FLiBe breeder. Total blanket volume is 342 m$^3$, so 1 g/m$^3$ of Pu-239 corresponds to 0.342 kg. At 100 wppm uranium in the beryllium, therefore, there is $\approx$1~kg of Pu-239 in the blanket after five years of full-power operation. These results indicate that it would take decades to amass 1 SQ of Pu-239 (8~kg) from naturally occurring uranium impurities in a single ARC-class FPP, although the amounts of Pu-239 produced are not trivial. This motivates careful attention to allowable beryllium impurity limits.}
\end{figure}

To perform this analysis, we used the same geometric model described in Sec.~\ref{subsec:modeloverview}, but implemented a new material definition that allowed the quantity of uranium in the blanket to be specified as a weight fraction of the beryllium in the FLiBe. We made use of a independent depletion calculation, which assumes that the flux spectrum in the blanket is negligibly perturbed by the evolution of its composition. Given the very small amount of fertile and fissile material involved, this is assumption is well justified. Results are shown in Fig.~\ref{fig:impurity_fissile_mass}, which plots the mass density of Pu-239 as a function of time in the blanket assuming continuous full power operation. We observe that it takes nearly a decade for the concentration of Pu-239 to reach its peak. For the liquid immersion blanket modeled here, with a volume of 342 cubic meters, this corresponds to a peak total Pu-239 mass of $\approx$ 3~kg for the 200~wppm impurity case, less than half of a significant quantity of WUM but possibly enough for a weapon \cite{Goddard2017}. 

We also observe that after a decade the quantity of Pu-239 in the blanket becomes approximately constant for the next two decades of operation, indicating that the rate of Pu-239 production and loss are approximately equal. Thus, while the total mass of plutonium in the blanket never exceeds a significant quantity, if plutonium was removed occasionally, a significant quantity could be accrued over time. Notably, it would take decades to produce 1 SQ of Pu-239 from a single ARC-class FPP based on this model (especially at uranium impurity levels $<$200~wppm, as would be expected from most naturally occurring beryllium). However, similar timescales have applied to other national weapons programs. It is also possible that 1 SQ could be amassed by collecting the outputs from multiple ARC-class FPPs, reducing the time needed to accrue a significant quantity. In general, uranium impurities in beryllium are not expected to pose an urgent breakout risk. At the same time, the amount of Pu-239 produced in the blanket from impurities is not necessarily trivial: at 100~wppm uranium impurity in the beryllium, a kilogram of Pu-239 will be amassed in the blanket at five years of full-power operation. In general, though, breakout risk posed by naturally occurring uranium impurities can be effectively mitigated by mandating lower allowable uranium impurity levels in the FLiBe.

\section{Discussion} 

Table \ref{tab:summary} reviews some of the key findings from Sec.~\ref{sec:results} and summarizes their implications on the breakout proliferation risk associated with the ARC-class FPP modeled in this work. In general, the results indicate that an SQ of high-purity WUM can be rapidly produced in the breeding blanket of an ARC-class FPP. In this section, we discuss how this might affect global security and the emerging fusion industry.

The feasibility of the proliferation scheme outlined here ultimately depends on the technical sophistication of the state attempting it. In general, states operating FPPs will probably have a baseline level of expertise that is more than adequate to initiate a program to produce WUM. Acquiring the raw fertile material (U or Th) in the needed ton quantities is straightforward. These elements are about as abundant as tin, and virtually every country has adequate resources for a weapons program. The periphery of sandstone aquifers and phosphate mines are especially rich sources. Iraq, for example, acquired the uranium for its nuclear-weapons program by re-milling the tailings from a phosphate mine \cite{Kemp2006}. The technology for large-scale salt cleanup may need to be custom built, but if the FPP comes with an online salt-purification system, that system might provide the required template. The weapons technology itself is eighty years old, demonstrably within reach of states like North Korea, and much simpler than an FPP to model and understand. In general, there are very few technical barriers to making nuclear weapons once the fissile material is in hand \cite{DOE1997}. 

The decision to proliferate is constrained primarily by politics, but when national leaders become motivated to acquire nuclear weapons, they tend to view the effort as essential to their continued existence as a nation \cite{Goheen1983,Thayer1995,Levite2002}. As such, even high-value assets, like FPPs, can be conscripted into service: of the 32 countries \cite{Levite2002} that have entertained nuclear-weapons programs, an estimated 70\% drew up plans to use their civilian nuclear-power infrastructure to jumpstart weapons production \cite{Kemp2019}. The potential to exploit peaceful energy technologies for weapons motivated the creation of the IAEA in 1957, but that agency lacks the resources to stop proliferation---its best hope is early detection, followed by a drawn-out international review of anomalies followed by a collective international action, as occurred most recently with Iran \cite{IAEA_Aims_Limitations}. This process is not fast. For example, the IAEA opened its formal investigation of Iran's nuclear program in the summer of 2002, but it took more than four years before the first United Nations resolution finding Iran in noncompliance was passed in December 2006. The weeks-to-months breakout timelines discussed here are much faster than the international community can typically respond. 

IAEA Safeguards are designed to confirm the non-diversion of 1~SQ of fissile material once every year from declared facilities that routinely process Special Fissionable or Source Materials \cite{IAEA_Statute}. Under normal operation, FPPs would not be routinely inspected because they are not expected to possess qualifying nuclear materials. Putting this legal issue aside, there is a practical problem with inspecting FPPs. Fission plants can be inspected effectively on an annual basis because the fissile material in them ($\gg\!1$~SQ) is stored in solid fuel bundles, which are quick to verify by counting. By contrast, the fissile-material breeding process described here is more akin to a bulk processing facility, such as occurs in reprocessing plants, where safeguards inspections need be very frequent or continuous.\footnote{In fact, the IAEA doesn't have the resources to meet its inspection goals for all bulk-process plants, and there are comparatively few such plants in the world. A fusion future would put enormous strain on the IAEA if a bulk-plant standard were needed.} If the IAEA were to adopt the goal of detecting proliferation at FPPs, rather than merely confirming non-diversion of declared material, inspections may be needed on timescales of the $t_\text{SQ}$ estimates identified above. 

In point of fact, however, IAEA Safeguards have not actually detected proliferation activities. Inspections are choreographed events known to the inspected state well in advance.\footnote{Special Inspections theoretically give the IAEA the power to inspect any sites with very limited warning, but these inspections have not been routinely used for fear of political repercussions \cite{IAEA1998, Heinonen2012}. The Additional Protocol allows Complementary Access at routinely inspected sites, with one day notice, but under present rules this option would not apply to FPPs because the Additional Protocol doesn't apply to sites that don't routinely process Special Fissionable or Source Materials per Article~XX of the IAEA Statute.} This limitation of Safeguards has led some countries to take matters into their own hands. Israel, for example, used military attacks in 1981 and 2007 to stop nuclear reactors from making weapons in Iraq and Syria, respectively; the United States went to war with Iraq notionally for this purpose; and the United States has used coercion in numerous other cases\footnote{Successfully for Taiwan and South Korea, and unsuccessfully for India, Pakistan, and North Korea.}. As such, actual proliferation prevention depends on the ability of powerful nations to detect weapons programs on their own, and on having adequate and timely options to reverse the program, be they diplomatic or military in nature. 

Nevertheless, the key risk for the proliferator is early detection that might lead to an effective intervention. The preparatory activities for fissile breeding can probably be carried out with virtually no risk of detection \cite{Kemp_2016}. We also showed that the excess heat from fission during the breeding process was small, and could easily be offset by slightly reducing the fusion power, making detection by thermal emission infeasible. If there is tamper-proof monitoring of the blanket by the international community, then its use for fissile material breeding would be readily detected. There may also be signatures that could be detected at the fence-line of the plant \cite{Kemp2019}. However, these do not ensure that the response will be timely or adequate. Because a kinetic military strike against an operating FPP could result in tremendous radiological release and subsequent health consequences, counterproliferation actions might be limited to political options, which depend entirely on the equities of the state doing the proliferating. In other words, there may be no {\it ex-post-facto}\/ technological fix once a state commences WUM production.

In view of these considerations, ARC-type FPPs might be regarded by governments as presenting nontrivial proliferation risks. Because the ability to stop proliferation might rest heavily on the identity of the proliferator, restrictions on FPP export may be an important element in preventing their eventual misuse.\footnote{Even so, political relationships are unpredictable. In the 1970s, the United States was an enthusiastic exporter of nuclear technology to Iran, only to regret that decision a decade later.} Designing FPPs to be more intrinsically resistant to proliferation may help abate these concerns. Li-6 enrichment can help extend the breakout timelines, giving more time for a response, although this can be partially overcome if domestic FLiBe production is possible. Exporting plants incapable of tritium self sufficiency would be a way to ensure a continued dependence on a supplier nation that could monitor the use of the plant; but it seems unlikely that FPP buyers will be enthusiastic to make multi-billion dollar investments that permanently tie their energy security to a supplier nation. It would also require that supplier nations operated FPPs that could produce enough excess tritium to keep other nations operating. This two-tier system ultimately requires buyers to go along with the restriction, and in the event there are multiple international suppliers, competition may lead other suppliers to undercut this scheme. 

\begin{table*}[h]
\caption{Key findings for both the U-238 $\rightarrow$ Pu-239 and Th-232 $\rightarrow$ U-233 breakout scenarios in an ARC-class FPP, and possible implications of the quantitative results on breakout proliferation risk.}
\label{tab:summary}
\scriptsize
\begin{tabular}{p{3.6cm} p{4cm} p{4cm} p{4cm}}
\toprule
\textbf{Consideration}	&	\textbf{U-238 $\rightarrow$ Pu-239} 	&	\textbf{Th-232$\rightarrow$U-233} 	&	\textbf{Summary} 	\\
\toprule
Amount of fertile mass required to breed 1~SQ in $<$~1~month	&	$\approx$ 20 metric tons of U-238 needed (approximately 1 m$^3$). Natural uranium is 99.3\% U-238. 	&	Requires $>$ 50 metric tons of Th-232 (beyond range considered here) because of 27 day intermediate daughter half-life.	&	Obtaining raw material not a major deterrent for U-238; possibly a deterrent for Th-232	\\
\hline
Amount of fertile mass required to breed 1~SQ  in $<$~2~months	&	$\approx$ 7 metric tons of U-238 needed ($<$ 0.5~m$^3$) 	&	$\approx$ 15 metric tons of Th-232 needed ($\approx$ 1.3~m$^3$). Natural thorium is  99.98\% Th-232.	&	Obtaining raw material not a major deterrent	\\
\hline
Amount of fertile mass required to breed 1~SQ in $<$~1~year (approx.\ IAEA inspection timeline)	&	$\approx$ 1 metric ton (calculated with empirical fit given in Sec.~\ref{subsec:tsq})	&	$\approx$~1.3 metric tons (calculated with empirical fit given in Sec. \ref{subsec:tsq}) 	&	Obtaining raw material not a major deterrent	\\
\hline
Addition of fertile mass and impact on tritium breeding	&	For fertile masses plotted, TBR does not drop below levels required for tritium self-sufficiency in any ARC-class FPP scenarios modeled in \cite{meschini2023modeling}. 	&	For fertile masses plotted, TBR does not drop below levels required for tritium self-sufficiency in any ARC-class FPP scenarios modeled in \cite{meschini2023modeling}. 	&	Addition of fertile mass not expected to affect tritium breeding to the point that plant is no longer able to operate. 	\\
\hline
Impact of excess heat due to fission in the blanket 	&	Increase of 10-15\% in total power handled by blanket 	&	Increase of 5--10\% in total power handled by blanket 	&	Not an issue to plant operation, unless heat exchange systems are designed with very small safety margins for excess power 	\\ 
\hline
Impact of excess heat due to radioactive decay of fertile and fissile material in the blanket 	&	Negligible perturbation to total power in blanket, reduces by order of magnitude $\sim$20 days after shutdown at $t_{\text{SQ}}$ 	&	Negligible perturbation to total power in blanket, reduces by order of magnitude $\sim$100 days after shutdown at $t_{\text{SQ}}$ 	&	Not an issue to plant operation, but may create a challenge for external reprocessing.	\\ 
\hline
Quality of fertile material produced at $t_{\text{SQ}}$	&	Purity above 99.0\% for all cases studied	&	Purity above 99.0\% for all cases studied, but substantial ($>$100 ppm) U-232 contamination &	For both Pu-239 and U-233 production schemes, resultant purity is well in excess of what is considered adequate for weapons-usable material, but U-233 is made less usable by U-232 impurity.	\\ 
\hline
Self-protection time 	&	Self-protection time is short ($\sim$1 day or less) &	 Self-protection time is short ($\sim$few hours) & The radiological hazard posed by fission products in the salt is unlikely to prevent the removal of the salt from the plant or greatly complicate its reprocessing, thus it does not impact the feasibility of fissile breeding. \\
\hline
Li-6 enrichment	&	Li-6 enrichment increases $t_{\text{SQ}}$ at all fertile masses considered. It also tends to increase TBR (except at very high Li-6 enrichment), decrease fission power in the blanket, decrease self-protection time, and decrease decay heat. 	&	Li-6 enrichment increases $t_{\text{SQ}}$ at all fertile masses considered. It also tends to increase TBR (except at very high Li-6 enrichment), decrease fission power in the blanket, decrease self-protection time, and decrease decay heat. 	&	Li-6 enrichment may be an effective way to add proliferation resistance by making it much harder to breed an SQ of WUM in between inspection periods. 	\\ 

\bottomrule    
\end{tabular}
\end{table*}

\section{Conclusion}

In this paper we have analyzed the risk of a fissile breeding breakout proliferation scenario in an ARC-class FPP by using the OpenMC Monte-Carlo neutronics code to perform a fully self-consistent time-dependent simulation of a simplified ARC-class blanket model. We examined the impact of two initial conditions: mass of fertile material in the breeding blanket and Li-6 enrichment of the breeding material, on six relevant breeding parameters: $t_{\text{SQ}}$, TBR, fission power, isotopic purity of bred WUM, self-protection time, and decay heat. We performed this analysis for two fissile breeding production schemes, U-238 $\rightarrow$ Pu-239 and Th-232 $\rightarrow$ U-233. We find that for all fertile mass inventories analyzed (5--50 metric tons) and a natural Li-6 enrichment (7.5\%), a significant quantity of WUM can be bred in less than six months of full power operation. This result indicates that ARC-class FPPs could pose a non-negligible proliferation risk if other mechanisms do not limit the feasibility of these scenarios.

We found that with a natural Li-6 enrichment of the blanket material, TBR, isotopic purity, fission power, self-protection time, and decay heat did not represent definitive limitations on the feasibility of this scenario. However we noted that U-233 bred from Th-232 is contaminated with a substantial amount of U-232, greatly increasing the radiological hazard posed by the final reprocessed fissile material, reducing its weapons usability.

We also varied the Li-6 enrichment of the breeding fluid and examined its impact on fissile breeding. Most consequently, we observed a substantial suppression of fissile breeding with increasing Li-6 enrichment, leading to a large increase in $t_{\text{SQ}}$. This alone provides a strong motivation for additional study of Li-6 enrichment as a tool for proliferation resistance. We also observe reductions in other relevant breeding parameters like fission power, self-protection time, and decay heat. TBR is sightly increased with Li-6 enrichment, as expected. The impact of introducing fertile material on TBR changes substantially over the enrichment interval studied, with lower Li-6 enrichments being more sensitive to the introduction of fertile material than higher enrichments. Total isotopic purity does vary with Li-6 enrichment, but not to any level that would impact its weapons usability. The concentration of U-232 in U-233 however does monotonically increase with Li-6 enrichment, further motivating the use of Li-6 enrichment for proliferation resistance.

We suggest that future work focus on the following:
\begin{itemize}
    \item Improving available workflows for self-consistent time-dependent fissile breeding calculations
    \item Use of more complete geometric models to better estimate TBR and parasitic neutron absorption in structural materials
    \item Exploration of detection and mitigation technologies
    \item Analysis of pathways for fertile material introduction / fissile material removal in various blanket types
\end{itemize}

In this work we have shown that, left unaddressed, ARC-class FPPs could pose a proliferation risk via fissile breeding. By analyzing and understanding this risk now, the fusion field has the opportunity to research and incorporate anti-proliferation measures into FPPs early in the design proces. For example, we have shown that the intrinsic proliferation resistance of D-T fusion reactors can be improved through Li-6 enrichment of the blanket material. 

While plans exist to build fusion pilot plants within the next ten years, it is unlikely that the issue of proliferation will become a barrier to fusion deployment until a global industry seeks to deploy many FPPs around the world. Early consideration of proliferation risks will work in service of that vision.  


\section{Acknowledgments}
The authors thank S. J. Frank for many useful discussions during the preparation of this paper, especially as it pertained to the review of prior work on this topic.  

\bibliographystyle{unsrt}
\bibliography{paper.bib}
\end{document}